\documentclass[aps,prl,reprint,superscriptaddress,longbibliography,nofootinbib]{revtex4-1}
\usepackage{lipsum}

\usepackage[utf8]{inputenc}

\usepackage{ tipa }
\usepackage{amsfonts}
\usepackage{amsmath}
\usepackage{tikz}
\usetikzlibrary{arrows,calc}
\usetikzlibrary{patterns,snakes}

\usepackage{ textcomp }

\usepackage{ upgreek }

\newcommand{\R}{\ensuremath{\mathbb R}}

\DeclareMathOperator{\Tr}{Tr}

\newcommand{\bea}{\begin{eqnarray}}
\newcommand{\eea}{\end{eqnarray}}
\newcommand{\Op}{{\mathcal O}}
\newcommand{\f}{{\mathfrak f}}

\begin{document}


\title{Krylov complexity in conformal field theory}


\author{Anatoly Dymarsky}
\affiliation{Department of Physics and Astronomy, \\ University of Kentucky, Lexington, KY 40506\\[2pt]}
\affiliation{Skolkovo Institute of Science and Technology, \\ Skolkovo Innovation Center, Moscow, Russia, 143026\\[2pt]}
\author{Michael Smolkin}
\affiliation{The Racah Institute of Physics, The Hebrew University of Jerusalem, \\ Jerusalem 91904, Israel \\}


\date{\today}
\begin{abstract}
Krylov complexity, or K-complexity for short, has recently emerged as a new probe of chaos in quantum systems. It is a measure of operator growth in Krylov space, which conjecturally bounds the operator growth measured by the out of time ordered correlator (OTOC). 
We study Krylov complexity in conformal field theories by considering arbitrary 2d CFTs, free field, and holographic models. We find that the bound on OTOC provided by Krylov complexity reduces to bound on chaos of Maldacena, Shenker, and Stanford.
 In all considered examples  
including free and rational CFTs Krylov complexity grows exponentially, in stark violation of the expectation that exponential growth signifies chaos. 

\end{abstract}


\maketitle

Quantum chaos and complexity play increasingly important role in understanding dynamical aspects of quantum field theory and quantum gravity. The notion of quantum chaos is difficult  to define and there are different complementary approaches. The conventional approach in the context of quantum many-body systems is rooted in spectral statistics, Eigenstate Thermalization Hypothesis (ETH), and absence of integrability \cite{d2016quantum}. In the context of field theory and large N models another well-studied signature of chaos is the behavior of the out of time ordered correlator (OTOC) \cite{maldacena2016bound}. These approaches focus on different aspects of quantum dynamics and usually apply to different systems. It is an outstanding problem to develop a uniform approach to chaos which would connect and unite them.  Dynamics of quantum operators in Krylov space has been recently proposed as a potential bridge connecting dynamics of OTOC with the conventional signatures of many-body chaos \cite{parker2019universal}. 

Krylov space is defined as the linear span of nested commutators $[H\dots,[H,{\cal O}]]$, where $H$ is the system's Hamiltonian and $\cal O$ is an operator in question.
Accordingly, time evolution $\Op(t)$ can be described as dynamics in Krylov space. Krylov complexity $K_\Op(t)$ defined below in \eqref{Kcomplexity} is a measure of operator size growth in {\it Krylov space}. For the chaotic systems it is expected to grow exponentially \cite{parker2019universal}, $K_\Op(t) \propto e^{\lambda_K t}$, the point we further elucidate below. For systems with finite-dimensional local Hilbert space, e.g.~SYK model \cite{PhysRevD.94.106002,rosenhaus2019introduction,trunin2020pedagogical}, it has been shown that at infinite temperature $\lambda_K$ bounds Lypanunov exponent governing exponential growth of OTOC 
\bea
\lambda\leq \lambda_K. \label{conjecture}
\eea 
This inequality conjecturally applies at finite temperature $\beta>0$. From one side connection of Krylov complexity to OTOC is not that surprising given that the latter measures spatial operator growth \cite{roberts2015localized}. From another side, dynamics in Krylov space is fully determined in terms of thermal 2pt function, as discussed below. Hence, the bound on OTOC in terms of $K_\Op(t)$ is the bound on thermal 4pt function in terms of thermal 2pt function. In this sense it is similar to the proposals of \cite{hartman2017upper} and also \cite{murthy2019bounds}, which derives the Maldacena-Shenker-Stanford (MSS) bound on chaos \cite{maldacena2016bound} 
\bea
\lambda\leq 2\pi/\beta \label{MSS}
\eea 
from the ETH. From the effective field theory point of view the 4pt function is independent from the 2pt one, hence such a bound could only be very general and  apply universally. One may not expect that a general theory would saturate the bound, casting doubt on the proposal that the exponent $\lambda_K$ that controls the growth of Krylov complexity is indicative of the Lyapunov exponent $\lambda$. Indeed, we will see that in case of CFT models the conjectural bound \eqref{conjecture} holds but reduces to MSS bound \eqref{MSS} such that $\lambda_K$ would remain finite even when  $\lambda$  
would approach zero or may not be well defined. 

There is another aspect of Krylov complexity which makes it an important topic of study in the context of quantum field theory and holographic correspondence. 
Krylov complexity is one of the family of q-complexities introduced in \cite{parker2019universal}. At the level of definition it is not related to circuit complexity, but a number of recent works \cite{barbon2019evolution,jian2020complexity,rabinovici2020operator} found qualitative agreement between the behavior of $K_\Op(t)$ with the behavior of circuit and holographic complexities \cite{brown2016holographic}. We further comment on possible similarity in the case of CFTs  in the conclusions.

To conclude the introductory part, we remark that  Krylov complexity, and dynamics in Krylov space in general, is fully specified by the properties of thermal 2pt function. Our results therefore should be seen in a broader context of studying thermal 2pt function in holographic settings with the goal of elucidating quantum gravity in the bulk \cite{nunez2003ads,fidkowski2004black,festuccia2006excursions,festuccia2009bohr,iliesiu2018conformal,alday2020holographic,karlsson2021thermalization,rodriguez2021correlation}.

To remind the reader, we briefly introduce main notions of Krylov space. More details can be found in \cite{parker2019universal,dymarsky2020quantum}. Starting from an operator $\Op$ one introduces iterative relation 
\bea
{\cal O}_{n+1}=[H,{\cal O}_n]-b_{n-1}^2 {\cal O}_{n-1}, \label{iterative}
\eea 
where positive real  Lanczos coefficients $b_n$ are uniquely fixed by the requirement that ${\cal O}_n$ are mutually orthogonal with respect to scalar product 
\bea
\Tr(e^{-\beta H/2}{\cal O}_n e^{-\beta H/2} {\cal O}_m) \propto \delta_{nm}.
\eea
Lanczos coefficients depend on the choice of the system Hamiltonian $H$, the operator ${\cal O}_0={\cal O}$, and inverse temperature $\beta$. Time evolution of the operator can be represented in terms of Krylov space, 
\bea
\Op(t)\equiv e^{i H t}\Op e^{-i Ht}=\sum_{n=0}^\infty \varphi(t)_n \Op_n,
\eea
where normalized ``wave-function'' $\varphi_n(t)$ satisfies discretized ``Schr$\ddot{\rm o}$dinger'' equation 
\bea
\label{Scho}
-i{d\varphi_n\over dt}=b_{n} \varphi_{n+1}+b_{n-1} \varphi_{n-1},
\eea
with the initial condition $\varphi_n(0)=\delta_{n,0}$. It describes hopping of a quantum-mechanical ``particle'' on a one-dimensional chain.
Krylov complexity is defined as the averaged value of an ``operator'' $\hat{n}$ measured in the ``state'' $\varphi$, where for convenience index $n$ is shifted by $1$, 
\bea
\label{Kcomplexity}
K_\Op(t)\equiv (\Op|\hat{n}|\Op)=1+\sum_{n=0}^\infty n |\varphi_n(t)|^2.
\eea
One can similarly define K-entropy \cite{barbon2019evolution}
\bea
\label{Kentropy}
S_\Op(t)\equiv -\sum_{n=0}^\infty |\varphi_n|^2 \ln |\varphi_n|^2.
\eea

Lanczos coefficients, and hence $K_\Op(t)$, are encoded in thermal  Wightman 2pt function 
\bea\nonumber
C_0(\tau)&=&\langle \Op(- i (\tau+\beta/2))\Op(0)\rangle_\beta \\ &\propto& \Tr( e^{-({\beta\over 2}-\tau) H} \Op e^{-({\beta\over 2}+\tau)H}\Op).
\eea
Precise relation evaluating $b_n^2$ in terms of $C_0$ and its derivatives is discussed in Supplemental Material. We only note here that $b_n^2$ do not change under multiplication of $C_0$ by an overall constant.

In full generality for a physical system with local interactions $C_0(\tau)$ is analytic in the vicinity of $\tau=0$. This implies that power spectrum 
\bea
f^2(\omega)=\int\, dt\, e^{i\omega t }C_0(i t) \label{powerspectrum}
\eea
decays at large $\omega$ at least exponentially, 
\bea
f^2(\omega) \sim e^{-\tau^* \omega},\quad \omega \rightarrow \infty, \label{exptail}
\eea 
where $\tau^*>0$ is the location of first singularity of $C_0(\tau)$ along the imaginary axis, if any. 
It was anticipated long ago that the high frequency behavior of $f^2(\omega)$ for a  local operator in many-body system can be used as a signature of chaos. In particular exponential behavior  \eqref{exptail} was proposed as a signature of chaos in {\it classical} systems in \cite{elsayed2014signatures}. An equivalent formulation in terms of the singularity of $C_0(\tau)$ was proposed as a signature of chaos for {\it quantum} many-body systems in \cite{avdoshkin2020euclidean} based on the rigorous bounds constraining the magnitude of $C_0(\tau)$ in the complex plane. A further step had been taken  in \cite{parker2019universal} who proposed the universal operator growth hypothesis: in generic, i.e.~chaotic quantum many-body systems Lanczos coefficients $b_n^2$ associated with a local $\Op$ exhibit maximal growth rate compatible with locality, 
\bea
b_n\approx \left({\pi \over 2\tau^*}\right) n +o(n),\qquad n\gg 1.  \label{UOGH}
\eea
This is stronger than the exponential behavior \eqref{exptail}, i.e.~it implies the latter, and reduces to it upon an additional assumption that the behavior of $b_n^2$ as a function of $n$ is sufficiently smooth for $n\rightarrow \infty$.   Modulo similar assumption of smoothness of $b_n^2$ ref.~\cite{parker2019universal} proved that in this case Krylov complexity grows exponentially as 
\bea
K_\Op(t) \propto e^{\lambda_K t},
\eea
where $\lambda_K=\pi/\tau^*$.

In field theory $C_0(\tau)$  necessarily has singularity at $\tau=\beta/2$, implying exponential decay of the power spectrum \eqref{exptail} with $\tau^*=\beta/2$.  
{\it Assuming} sufficient smoothness of $b_n^2$, one immediately arrives at  \eqref{UOGH} \cite{lubinsky1993update,basor2001determinants} (also see \cite{parker2019universal,avdoshkin2020euclidean}), and exponential growth of Krylov complexity with $\lambda_K=2\pi/\beta$. Hence the conjectural bound on OTOC \eqref{conjecture} reduces to the MSS bound \eqref{MSS}. This logic applies to {\it any} quantum field theory, including free, integrable or rational CFT models. Similarly, one can conclude that for field theories universal operator growth hypothesis \eqref{UOGH} trivially holds, but the exponential behavior of Krylov complexity can not be regarded as an indication of chaos. We stress, these conclusions are premature as one needs to justify the smoothness assumption by e.g.~evaluating $b_n^2$ explicitly.
 Without this assumption asymptotic behavior of $b_n^2$ is {\it not} determined by the high frequency tail of $f^2(\omega)$, or the singularity of $C_0(\tau)$, as is shown explicitly by a counterexample in \cite{avdoshkin2020euclidean}. We justify the smoothness assumption  by considering several different CFT models and evaluating Lanczos coefficients.

i)  In case of {\bf 2d CFTs} thermal 2pt function of primary operators $\Op$ is fixed  by conformal invariance
\bea
C_0={1\over \cos(\pi \tau /\beta)^{2\Delta}}, \label{2dCFT}
\eea 
where $\Delta$ is the dimension of $\Op$. This $C_0$ has been thoroughly analyzed in \cite{parker2019universal} in the context of SYK model. In particular they found $b_n^2=(n+1)(n+2\Delta)(\pi/\beta)^2$ and $K_\Op(t)=1+2\Delta \sinh^2(\pi t/\beta)$. In other words $b_n^2$ dependence on $n$ is smooth and Krylov complexity grows exponentially with $\lambda_K=2\pi/\beta$.

ii)  In case of {\bf free massless scalar in $d$ dimensions}, as well as {\bf Generalized Free Field} of conformal dimension $\Delta$  \cite{alday2020holographic}, thermal 2pt function is given by, 
\bea
\label{GFF}
C_0=c_d\left(\zeta(2\Delta, 1/2+\tau/\beta)+\zeta(2\Delta, 1/2-\tau/\beta)\right).\ \
\eea
Coefficient $c_d$ ensures canonical normalization in case of free massless scalar and is not important in what follows.  In the latter case $\Delta=d/2-1$.

\begin{figure}[t]
\includegraphics[width=0.5\textwidth]{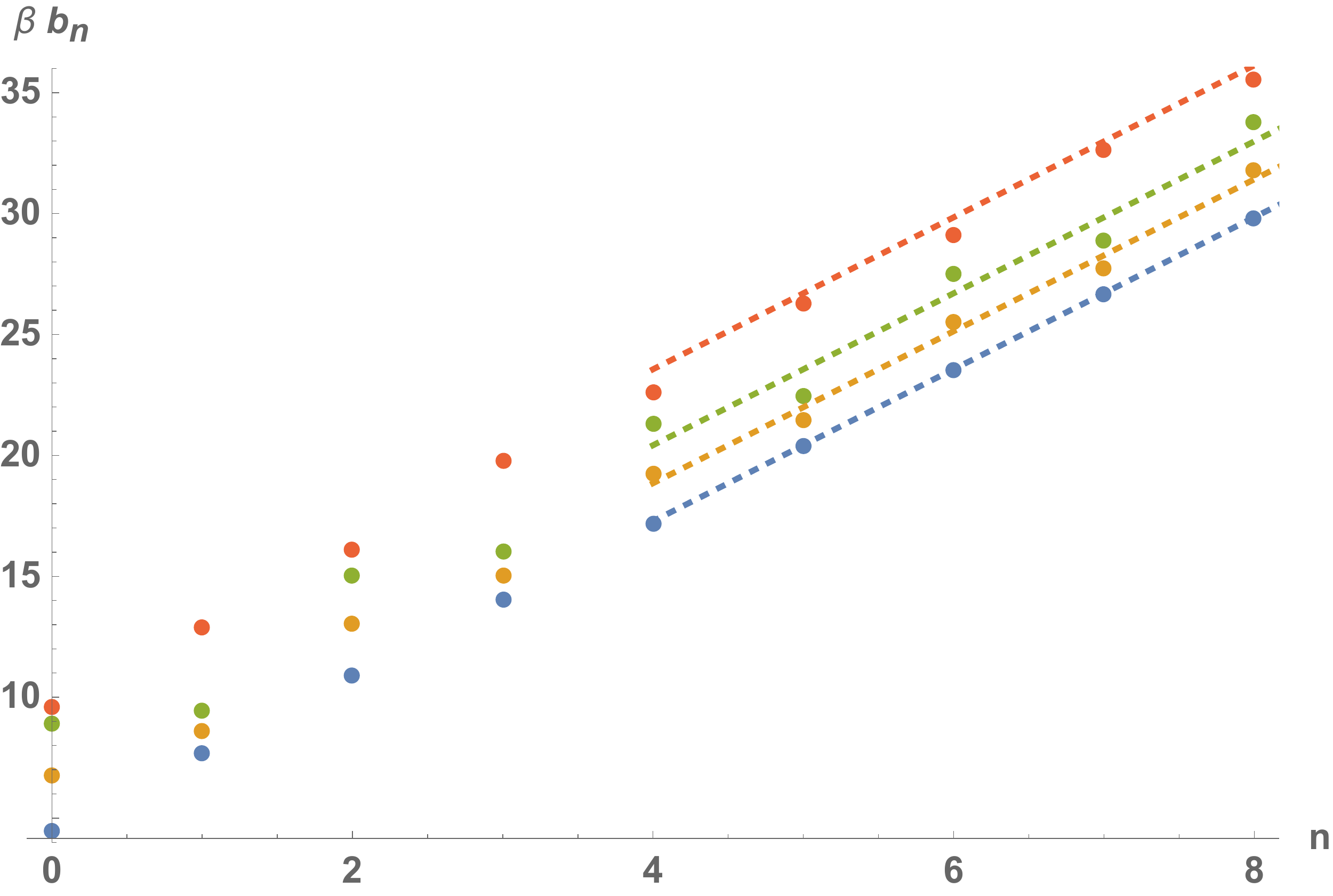}
\caption{Lanczos coefficients $b_n$ for free massless scalar $\phi$ in $d=4$  ($\Delta=1$, blue),
$d=5$ ($\Delta=3/2$, orange), $d=6$ ($\Delta=2$, green) dimensions, and for the composite operator $\phi^2$ in $d=5$ dimensions ($\Delta=3$, red); dashed lines of the appropriate color show asymptotic behavior  of $b_n$ as given by \eqref{asympt}.}
\label{fig:bn}
\end{figure}
For \eqref{GFF} with general $\Delta$ explicit expression for Lanczos coefficients is not known. In the special case of $d=4$, $C_0$ reduces to \eqref{2dCFT} with $\Delta=1$, and the rest applies. For $d=6$, $\Delta=2$, and Lanczos coefficients can be evaluated using connection to integrable Toda hierarchy \cite{dymarsky2020quantum},  yielding (see Supplemental Material)
\bea
\label{d=8}
b_n^2&=&\left({\pi/ \beta}\right)^2(n+2) (n+3) {g_{n-1} g_{n+1}\over g_n^2},\\
g_n&=& H_{n+2}+(-1)^{n+1} \Phi(-1,1,3+n)+\ln(2). \nonumber
\eea
\begin{figure}[t]
\vspace{0.2cm}
\includegraphics[width=0.5\textwidth]{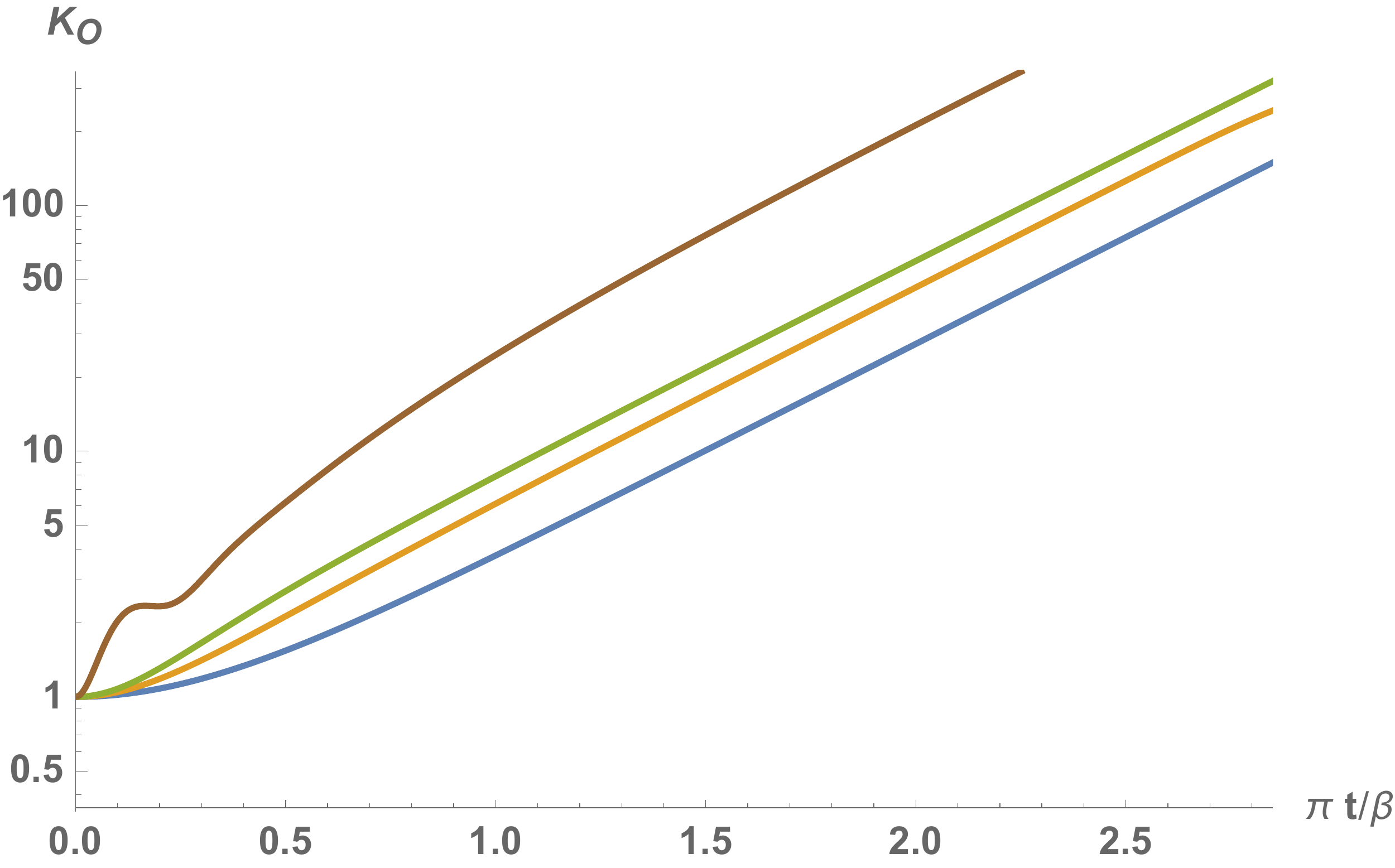}
\caption{Krylov complexity $K_\Op$ shown in logarithmic scale for free scalar in $d=4$  (blue),
$d=5$ (orange), $d=6$ (green) dimensions and for Generalized Free Field with $\Delta=10$ (brown). Blue curve is known analytically, $\ln(1+2\sinh^2(\pi t/\beta))$. All four curves exhibit an apparent linear growth of $\ln K_\Op \propto 2\pi t/\beta$ at late times.}
\label{fig:KO}
\vspace{-0.2cm}
\end{figure}
Here $H_n$ is the harmonic number and $\Phi$ is Lerch transcendent.
In this case $b_n^2$ demonstrate ``staggering'' or ``dimerization'' -- the sequences of $b_n^2$ for even and odd $n$ can be combined into two families, each approximately described by smooth functions $b_n=h_n+(-1)^n \tilde{h}_n$, where $h_n \approx (\pi/2\tau^*) n +o(n)$ for $n\gg 1$. This is shown in Fig.~\ref{fig:bn}. Such a behavior was analyzed in \cite{yates2020lifetime,yates2020dynamics}, where it was shown that 
for smooth functions $h_n,\tilde{h}_n$ in the large $n$ region ``Schr$\ddot{\rm{o}}$dinger equation'' \eqref{Scho} reduces to continuous Dirac equation with the space-dependent mass. In the case when asymptotically
$\tilde{h}_n \rightarrow 0$, mass eventually approaches zero for large $x$, describing propagation of a quantum ``particle'' with the speed of light $x(t)\sim t$  with respect to an auxiliary spatial continuous coordinate  $x$ which is related to $n$ via \cite{yates2020dynamics}
\bea
n\propto (e^{(2\pi/\beta) x} -1).
\eea
From this follows that for late times Krylov complexity will grow exponentially 
\bea
K_\Op(t)\approx e^{2\pi/\beta(t-t_0)} \label{Kexp}
\eea where $t_0$ is the characteristic time ``quantum particle'' described by $\varphi_n(t)$ will spend near the edge of the Krylov space $n\sim O(1)$. From the analytic expression for $K_\Op$ in case of 2d CFTs we conclude that $t_0$ is growing negative for large $\Delta$, $t_0\sim -\ln \Delta$.
The only scenario to avoid exponential growth of $K_\Op$ with $t$ is for $\varphi_n(t)$ to be localized  near the edge $n \sim O(1)$, which would presumably require erratic behavior of $b_n$ for small $n$.

Numerical simulation of $K_\Op$ for massless scalar in $d=6$ shown in Fig.~\ref{fig:KO}  confirms exponential behavior \eqref{Kexp} with  $t_0$ of order one. Thus, despite ``staggering'' Krylov complexity for free massless scalar in $d=6$ behaves qualitatively similar to $d=4$ case.

Next we numerically plot Lanczos coefficients for free scalar in $d=5$ with $\Delta=3/2$, see Fig.~\ref{fig:bn}. Similarly to $d=6$, $b_n$'s  exhibit staggering, which does not affect asymptotic exponential behavior of $K_\Op$, see Fig.~\ref{fig:KO}.

To analyze general case \eqref{GFF} with $\Delta\gg 1$ we can approximate $C_0$ with an exponential precision by 
\bea
C_0 \propto {1\over (\beta+2\tau)^{2\Delta}}+{1\over (\beta-2\tau)^{2\Delta}}.
\eea
By employing $1/\Delta$ expansion we find for small $n$
\begin{widetext}
\bea
\label{smalln}
\beta^2 b_n^2=
\left\{ 
\begin{array}{lc}
16 \Delta^2+  8(1 + 3 n)\Delta+ {19 n^2}/{2}+7 n +O(n^3)/\Delta+\dots & {\rm for}\ n\ {\rm even},\\
 16 (1 + n) \Delta+ 2 (n+1) (5 n+1)+O(n^3)/\Delta+\dots &  {\rm for}\  n\ {\rm odd}.
\end{array}
\right.
\eea
\end{widetext}
Thus, staggering grows with $\Delta$, but $n$ dependence of $b_n$ for odd and even $n$ remain smooth. 

\begin{figure}[b]
\includegraphics[width=0.24\textwidth]{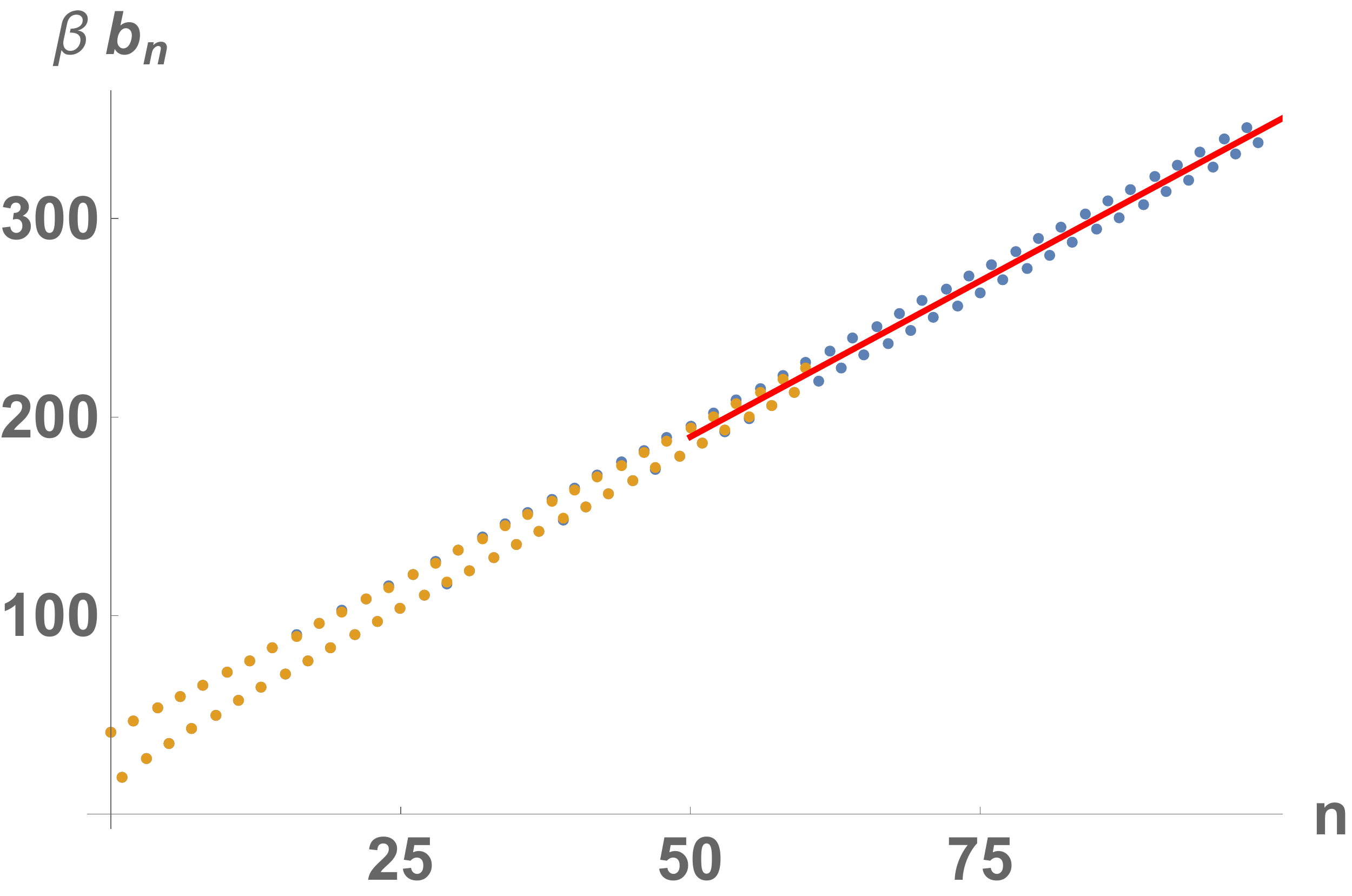}\includegraphics[width=0.24\textwidth]{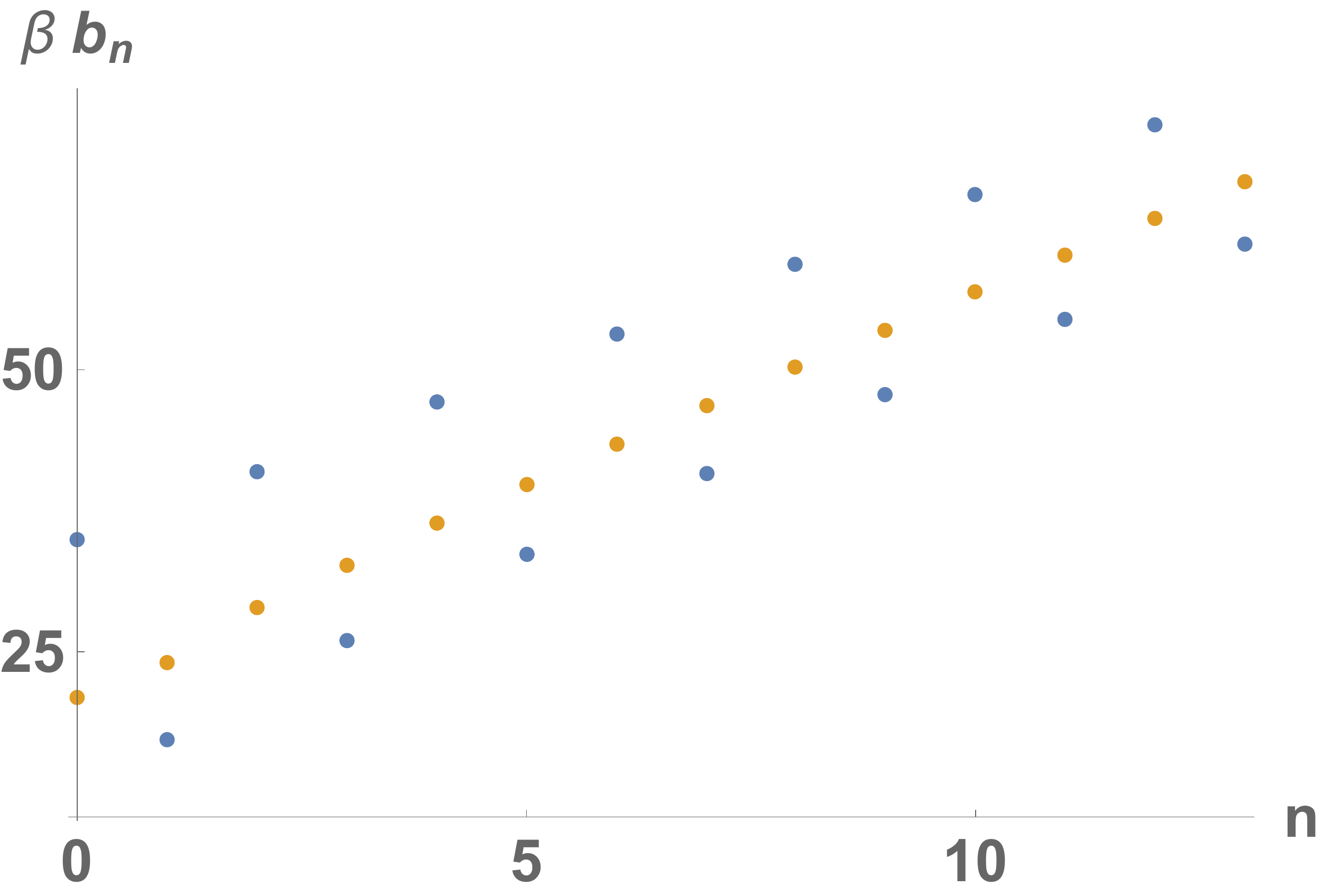}
\caption{Left panel. Lanczos coefficients $b_n$ for Generalized Free Field \eqref{GFF} with $\Delta=10$ (blue) vs approximation for small $n$ \eqref{smalln} (orange) and asymptotic behavior for large $n$ \eqref{asympt} (red line). Right panel. Lanczos coefficients $b_n$ for Generalized Free Field \eqref{GFF} of dimension $\Delta=8.5$ (blue) and for  holographic  operator $O=\int d^3 x\, \Op$ of effective dimension $\Delta=8.5$, while $\Op$ has dimension $\Delta=10$  (orange). The same effective dimension means both sequences have the same asymptotic behavior $b_n\approx \pi (n+9)$.}
\label{fig:bnGFF}
\label{fig:GFF-hol}
\end{figure}

For large $n$ pole structure of $C_0$ suggests, see Supplemental Material, 
\bea
\beta\, b_n\approx \pi(n+\Delta+1/2). \label{asympt}
\eea
These approximations accurately describe $b_n$ for small and large $n$ correspondingly, as is shown in the left panel of Fig.~\ref{fig:bnGFF}.
Numerical simulation of $K_\Op(t)$ for $\Delta=10$ shown in Fig.~\ref{fig:KO} confirms exponential behavior with $\lambda_K=2\pi/\beta$ and $t_0$ of order $-\ln\Delta$. In other words staggering, exhibited by $b_n$ in case o free scalar field, which grows with $\Delta$, is not affecting dynamics at late times -- $K_\Op$ grows exponentially with the exponent $\lambda_K=2\pi/\beta$, although dynamics at early times becomes more complicated.  

Finally, we discuss composite operators $\Op^m$ for some integer $m$. By Wick theorem Wightman function simply becomes $C_0\rightarrow C_0^m$ with an unimportant overall coefficient. In the case of 2d CFT or free massless scalar in $d=4$ we again obtain $C_0$ of the form \eqref{2dCFT}. In other cases Lanczos coefficients should be calculated numerically. We plot $b_n$ for $\Op=\phi^2$ in free massless scalar theory in $d=5$  in Fig.~\ref{fig:bn}.

iii) In case of {\bf free fermions in $d$ dimensions}, 
\bea
 C_\psi(\tau) &=& r_d  
 \left(  \zeta\big(2\Delta,{1\over 4}-{\tau\over 2 \beta}\big) + \zeta\big(2\Delta,{1\over 4}+{\tau\over 2\beta}\big) \right.
 \nonumber \\
 &-&\left. \zeta\big(2\Delta,{3\over 4}-{\tau\over 2\beta}\big) - \zeta\big(2\Delta,{3\over 4}+{\tau\over 2\beta}\big)  \right), \label{fermion}
\eea
where dimension of free fermion is $\Delta=(d-1)/2$. 
We notice that Lanczos coefficients for free fermion in dimension $d$ are very close to those for the free boson of the same conformal dimension $\Delta$, i.e.~in dimension $d+1$. The same applies for $b_n$ for the composite operators $\bar \psi \psi$ and $\phi^2$. Corresponding comparison is delegated to Supplemental Material.

iv) In case of {\bf holographic CFT} thermal two-point function can be calculated  by solving wave equation  in the bulk \cite{festuccia2006excursions,festuccia2009bohr}.
We perform this numerically in Supplemental Material to find that $b_n$ smoothly depend on $n$. This is shown in the right panel of Fig.~\ref{fig:GFF-hol} where we superimposed $b_n$ for the holographic model with Lanczos coefficients for the Generalized Free Field of the same effective dimension, determined by the singularity of $C_0$ near $\tau\rightarrow \beta/2$.
Smooth behavior  perfectly matches the expectation that for holographic theories exhibiting maximal chaos, $\lambda=2\pi/\beta$, growth of Krylov complexity also must be governed by the same exponent.

\begin{figure}[b]
\includegraphics[width=0.5\textwidth]{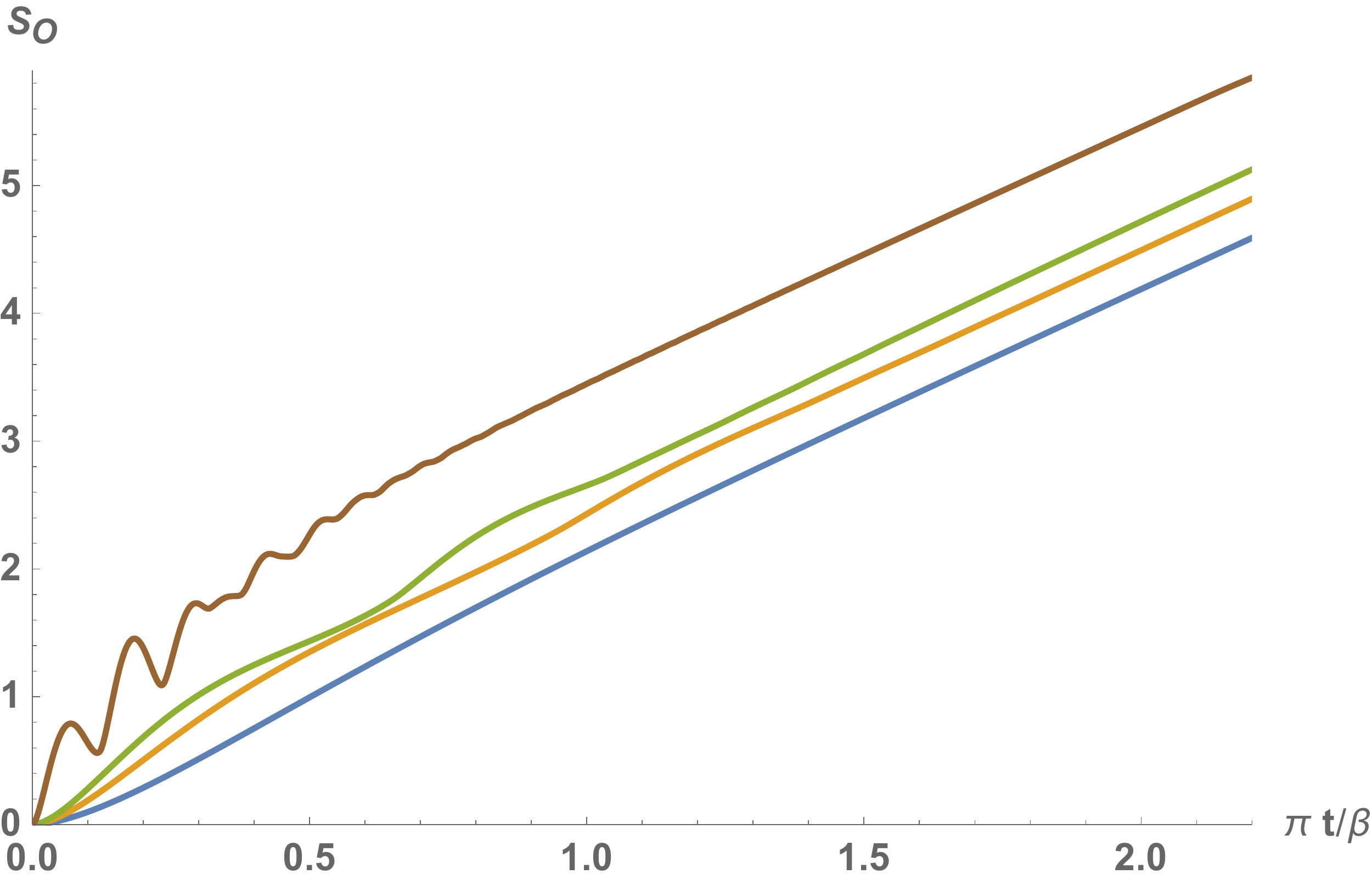}
\caption{K-entropy $S_\Op$ shown in logarithmic scale for free scalar in $d=4$  (blue),
$d=5$ (orange), $d=6$ (green) dimensions and for Generalized Free Field with $\Delta=10$ (brown).  All four curves exhibit an apparent linear growth  at late times.}
\label{fig:SO}
\end{figure}

Besides Krylov complexity we numerically plot growth of Krylov entropy \eqref{Kentropy} for several different models, shown in Fig.~\ref{fig:SO}. In all cases it exhibits linear behavior for late $t$, confirming scrambling of $\Op$  in Krylov space. We conclude that only early time dynamics is sensitive to pecularities of the model, while at late times dynamics in Krylov space exhibits remarkable universality.

\noindent
{\bf Conclusions.} In this paper we studied Lanczos coefficients and operator growth in Krylov space for local operators in various CFT models. For some models $b_n$ were calculated analytically, while for others we had to resort to numerical analysis. We also found  asymptotic behavior of $b_n$ for large $n$ \eqref{asympt}. One of the main goals was to study if Krylov complexity is sensitive to the underlying chaos. A general argument presented in the introduction dictates that so far asymptotic behavior of $b_n$ as a function of $n$ is sufficiently smooth, Lanczos coefficients exhibit universal operator growth hypothesis \eqref{UOGH} and Krylov complexity grows exponentially \eqref{Kexp}. The only possible caveat is the possibility that for large $n$ different subsequences of $b_n$ would have different asymptotic, for example $b_n$ for even and odd $n$ would grow as $n^a$ with different $a_{\rm even}\neq a_{odd}$. Another hypothetical possibility, which will not affect  \eqref{UOGH} but may affect \eqref{Kexp}, is that erratic behavior of $b_n$ for small $n$ will cause approximate or complete localization of the operator ``wave-function'' $\varphi_n$, leading to large or infinite $t_0$. We did not see any behavior of this sort in any model we considered, including arbitrary 2d CFTs, free bosons and fermions, composite operators, generalized free field of arbitrary dimension, and a holographic model in $d=4$. On the contrary we observed linear growth of $b_n$ at large $n$ in full agreement with \eqref{asympt} and exponential growth of Krylov complexity with $\lambda_K=2\pi/\beta$. In other words for considered models universal operator growth hypothesis of \cite{parker2019universal} trivially holds, and the conjectural bound \eqref{conjecture} on of OTOC at finite temperature in terms of growth of Krylov complexity reduces to MSS bound \cite{maldacena2016bound}. At the same time exponential growth of $K_\Op$ is not a signature of chaos as it grows with the same exponent $\lambda_K=2\pi/\beta$ for maximally chaotic holographic CFTs as well as for rational 2d CFTs and free field theories, for which Lypanunov exponent may not be even properly defined \cite{caputa2016scrambling,fan2018out,kudler2020conformal}. It would be interesting to extend our analysis for massive an interacting models, especially those exhibiting non-maximal chaos \cite{stanford2016many,murugan2017more,steinberg2019thermalization,mezei2020chaos}. Nevertheless since a continuous deformation can not change asymptotic behavior of $b_n$ we provisionally conclude our results will remain valid in the case of general interacting quantum field theory. 

It is also instructive to compare behavior of Krylov complexity with various notions of circuit and holographic complexities. While the latter are defined for states and the former is a measure of operator growth, an analysis of \cite{barbon2019evolution,jian2020complexity,rabinovici2020operator} in the context of SYK-type models revealed some qualitative similarities. In this spirit we notice that $K_\Op$ exhibits essentially the same behavior for both free and holographic theories, similarly to complexity action proposal, as discussed in \cite{jefferson2017circuit,PhysRevLett.120.121602}. There are also bulk complexity proposals specific for conformal theories \cite{caputa2017liouville,caputa2019quantum}, which exhibit robust universality due to extended symmetry. 
To complete the comparison, it would be important to go beyond thermodynamic limit by placing CFT on a compact background, e.g.~${\mathbb S}^{d-1}\times \R$. In this case one may hope to study qualitative behavior of $K_\Op$ beyond scrambling time $t\sim \ln S$, when the exponential complexity growth should become linear.

\acknowledgments
We thank Pawe\l\,   Caputa, Mark Mezei and Alexander Zhiboedov for discussions. This work is supported by the BSF grant 2016186. 

\newpage

\section{Supplemental Material}
\subsection{Lanczos coefficients from the thermal 2pt function}
Recursion method is closely related to integrable Toda chain \cite{dymarsky2020quantum}. In particular thermal 2pt function $C_0(\tau)$ should be understood as the tau-function of Toda hierarchy, $\tau_0\equiv C_0$. Other tau-functions $\tau_n$ are related to $C_0$ as follows.  One introduces $(n+1)\times (n+1)$, $n\geq 0$, Hankel matrix of derivatives 
\bea
{\mathcal M}^{(n)}_{ij}=C_0^{(i+j)}(\tau),
\eea
where $C_0^{(k)}(\tau)$ stands for $k$-th derivative of $C_0$. Then 
\bea
\label{Hankel}
\tau_n(\tau)={\rm det}\, {\mathcal M}^{(n)}.
\eea
Tau functions automatically satisfy Hirota bilinear relation
\bea
\label{Hirota}
\tau_n \ddot{\tau}_n-\dot{\tau_n}^2=\tau_{n+1}\tau_{n-1}, \quad \tau_{-1}\equiv 1.
\eea
At this point we can introduce $q_n$ via $\tau_n={\rm exp}(\sum_{0\leq k\leq n} q_n)$, such that  $C_0(\tau)=e^{q_0(\tau)}$. Functions $q_n$ satisfy Toda chain equations of motion. Lanczos coefficients are 
\bea
b_n^2=e^{q_{n+1}-q_n}={\tau_{n+1}\tau_{n-1}\over \tau_n^2}. \label{Lt}
\eea
Defined this way $b_n^2$ are functions of $\tau$. To evaluate Lanczos coefficients from \eqref{iterative} we need to take $\tau=0$ in \eqref{Lt}. This prescription is equivalent to evaluation of $b_n^2$ from the moments of $f^2(\omega)$ \eqref{powerspectrum} described in \cite{parker2019universal}.

There is an explicit family of solutions with the asymptotic behavior $b_n^2\propto n^2$ \cite{dymarsky2020quantum}
\bea
\nonumber
\tau_n(\tau)&=&{G(n+2)G(n+1+2\Delta)\over G(2\Delta)\Gamma(2\Delta)^{n+1}} {(\pi/\beta)^{n(n+1)}\over \cos(\pi  \tau/\beta)^{(n+2\Delta)(n+1)}},\\
\nonumber
q_n(\tau)&=&2n\ln(\pi/\beta)-(2n+2\Delta)\ln(\beta\cos(\pi \tau/\beta))+ \\ 
\label{exs}
\qquad && \ln(n! \Gamma(n+2\Delta)),\\
\nonumber
a_n(\tau)&=&(2n+2\Delta)(\pi/\beta) \tan (\pi \tau/\beta),\\
b^2_n(\tau)&=&{ (n+2\Delta)(n+1) (\pi/\beta)^2 \over \cos^{2}(\pi \tau /\beta)}, \nonumber
\eea 
where $G(x)$ is the Barnes  gamma function. This is the solution which appears in the context of $C_0$ for the 2d CFTs \eqref{2dCFT}.

\subsection{Free massless scalar in $d$ dimensions}
For convenience we introduce $\tilde{\tau}=\tau/\beta$. Thermal 2pt function in coordinate space is given by an integral of Matsubara propagator, 
\bea
\nonumber
 C_\phi(\tilde\tau,d) ={\beta^{2-d} \over (4\pi)^{d-1\over 2} \Gamma\big({d-1\over 2}\big)} \int_{0}^\infty dy \, y^{d-3}  
 {\cosh(y\tilde \tau)\over \sinh\big({y\over 2}\big)}.
\eea
The integral over $y$ can be evaluated yielding \eqref{GFF} with $2\Delta=d-2$ and 
\bea
c_d= \beta^{2-d} {\Gamma\big({d-2\over 2}\big)  \over 4\pi^{d\over 2}}.
\eea
Numerically Lanczos coefficients for \eqref{GFF} can be evaluated from (\ref{Hankel},\ref{Lt}) using
\bea
C_\phi^{(2n)}(0)=c_d  {2(2^{2\Delta+2n}-1)\beta^{-2n}\over  \Gamma\big(2\Delta\big)}
 \Gamma(2\Delta+2n)\zeta(2\Delta+2n).\nonumber \\
\eea

Up to an overall coefficient for $d=4$ \eqref{GFF} reduces to \eqref{2dCFT} with $\Delta=1$.
It follows from the integral representation above that for integer $d$ 
\bea
 C_\phi(\tilde\tau,d)\propto {d^2\over d\tilde \tau^2} C_\phi(\tilde\tau,d-2),
\eea
and therefore for $d=6$
\bea
\label{d6}
C_0(\tau) \propto {d^2 \over d\tau^2} {1\over \cos^2(\pi \tau/\beta)}.
\eea
To find $b_n^2$ we look for the tau-functions of the form 
\bea
\tau_n&=&{(\pi/\beta)^{\alpha_n} p_n(\cos^2(\pi \tau/\beta))\over \cos(\pi \tau/\beta )^{a_n}},\quad n\geq 1,\\
a_n&=&(n+a)(n+1),\qquad \alpha_n=(n+\alpha)(n+1), \nonumber
\eea
where $p_n(y)$ is a polynomial of degree $n+1$ and $p_{-1}=1$ such that  $\tau_{-1}=1$.
 Then Hirota equations \eqref{Hirota} become iterative equations for the polynomials 
\begin{widetext}
\bea
p_{n+1}={a_n p_n^2-4(1-y) y^2 (p_n')^2+2 y p_n((1-2y)p_n'+2(1-y)y p_n'')\over p_{n-1}}.\label{iter}
\eea
\end{widetext}
To match with \eqref{d6} we take $a=4$, $\alpha=2$ (this value is in fact arbitrary) and $p_0=1-2y/3$. Then 
\bea
p_n(y)=\prod_{k=0}^{n-1} a_k^{n-k} \sum_{k=0}^{n+1} {(-4)^k\over (1+k)^2} {\Gamma(2+n)\Gamma(4+n+k) y^k\over \Gamma (4+n) \Gamma(2+2k)\Gamma (2+n-k)}.\nonumber 
\eea
This can be evaluated for $y=1$ which corresponds to $\tau=0$, 
\bea
p_n(1)={H_{n+2}+\ln(2)+(-1)^{n+1}\Phi(-1,1,n+3)  \over (n+2)(n+3)}, \nonumber
\eea
where $H_n=\sum_{k=1}^n k^{-1}$ is the harmonic number and $\Phi(z,s,\alpha)=\sum_{k=0}^\infty z^k/(k+\alpha)^s$ is the Lerch transcendent.
From here we obtain \eqref{d=8}.

The same logic can be applied to $d=8$, in which case $a=6$, and $p_0=1 - y + 2 y^2/15$. Iterative relation \eqref{iter} gives
\bea
p_n(y)=\prod_{k=0}^{n-1} a_k^{n-k} \left(1-{(n+1)(n+6)\over 6}y+O(y^2)\right), \nonumber
\eea
but we were not able to find closed form analytic expression.

\subsection{Pole structure of $C_0$ controls asymptote of $b_n$}
Under the assumption that the $n$-dependence of $b_n$ is smooth, at least for large $n\gg 1$, the asymptote of $b_n$ is controlled by the poles of $C_0$, or equivalently high frequency behavior of $f^2(\omega)$.  For an operator of dimension $\Delta$ CFT correlator $C_0(\tau)$ would have a pole singularity $\propto (\tau-\tau^*)^{-2\Delta}$ where $\tau^*=\beta/2$, implying asymptotic behavior
\bea
\label{asymptotef2}
f^2(\omega) \propto e^{-\omega \tau^*} \omega^{2\Delta-1}
\eea
for large $\omega$. Using saddle point approximation we can estimate the moments
\bea
\label{M2k}
M_{2k}\equiv {\int_{\-\infty}^\infty d\omega f^2(\omega) \omega^{2k}\over \int_{\-\infty}^\infty d\omega f^2(\omega)} \approx 
{\left({2\tilde{k}\over e \tau^* }\right)^{2\tilde{k}}\over \left({2\Delta-1\over e \tau^* }\right)^{2\Delta-1}},
\eea
where $\tilde{k}=k+\Delta-1/2$.
Strictly speaking  for validity we need to require $\Delta \rightarrow \infty$. In the case when $\Delta$ is of order one the expression above is valid only in the sense of $k$-dependence in the limit of large $k$. Assuming smooth $k$-dependence of $b_k$  we will approximate it by 
\bea
\beta\, b_k\approx {\mathfrak a}\, k+{\mathfrak b},\quad k\gg 1.
\eea
It is tempting to rewrite it as $\beta\, b_k={\mathfrak a}\, \tilde{k}$, and identify ${\mathfrak b/\mathfrak a}$ with  $\Delta-1/2$. To justify that we will use the formalism of integral over Dyck paths which evaluates $M_{2k}$ in terms of $b_n$'s developed in  \cite{avdoshkin2020euclidean}. At the level of quasiclassical approximation, which gives leading contribution in the limit of large $k$, the moments are given by 
\bea
M_{2k} \approx e^{S},
\eea
where $S$ is the on-shell value of action 
\bea\nonumber
S[f(t)]=2k\int_0^1 dt \left(-p \ln p-(1-p)\ln(1-p)+b(2k \f)\right),\\
\label{S}
\eea
where $p\equiv (1+\f')/2$, Lanczos coefficients $b_n$ are described by the smooth function $b(n)$ and ${\mathfrak f}(t)$ satisfies boundary conditions $f(0)=f(1)=0$.
For $b(2k \f)= 2k \f {\mathfrak a}+{\mathfrak b}$ equation of motion reads
\bea
{\f''\over (\f')^2-1}={2k \mathfrak a\over 2k \mathfrak a \f+\mathfrak b}
\eea
with the solution 
\bea
\f(t)={\sin((\pi -2r)t+r)\over (\pi -2r)}-{\mathfrak b\over 2k \mathfrak a}, \label{f}
\eea
where $r$ is defined from the equation 
\bea
{\mathfrak b\over 2k \mathfrak a}={\sin(r)\over\pi -2r}.
\eea
When $k$ goes to infinity this gives the asymptote $r \approx {\mathfrak b \pi /2k \mathfrak a}$.
Plugging the solution \eqref{f} back into action \eqref{S} yields
\bea
S=2k \ln \left({4 k \mathfrak a\over e(\pi -2r)}\right)+2 {\mathfrak b\over \mathfrak a}\ln\cot(r/2).
\eea
Taking large $k$ limit we finally arrive at
\bea
M_{2k} \approx \left({4 k \mathfrak a\over e\pi}\right)^{2k}  e^{2\mathfrak b\over \mathfrak a} \left({4 k \mathfrak a\over e \mathfrak b}\right)^{2\mathfrak b\over \mathfrak a}.
\eea
Comparing with \eqref{M2k} $k$-dependence we first obtain the result well-appreciated in the literature, 
\bea
\mathfrak a={\pi\over 2\tau^*},
\eea
and then also
\bea
{\mathfrak b\over \mathfrak a}=\Delta-1/2.
\eea
In this subsection we used the formalism of \cite{avdoshkin2020euclidean} which numerates $b_n$ starting from $n=1$. In the main text index $n$ starts from zero, yielding a shift by one in \eqref{asympt}.

\subsection{Free massless fermion in $d$ dimensions}
\begin{figure}[t]
\includegraphics[width=0.5\textwidth]{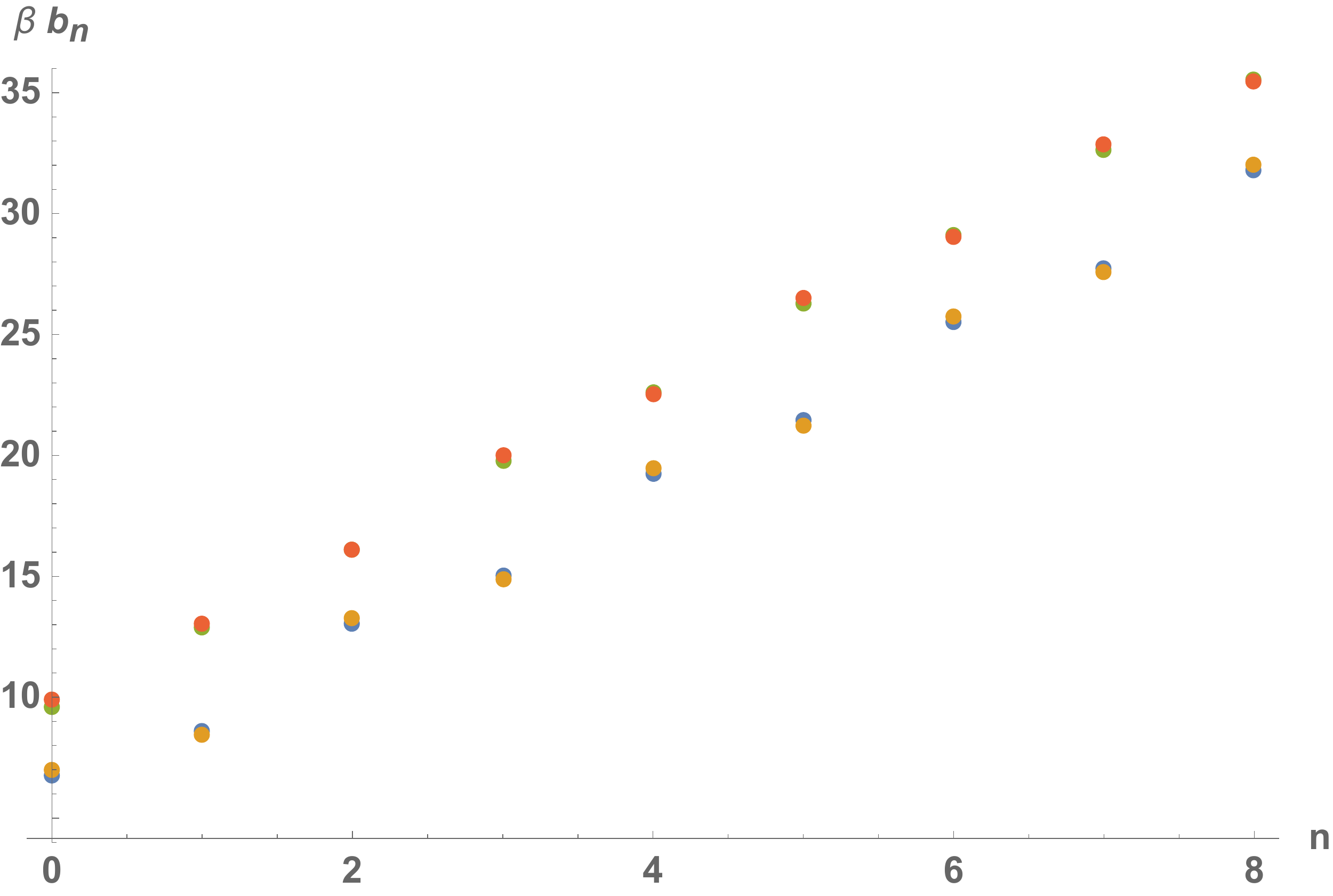}
\caption{Lanczos coefficients for free scalar \eqref{GFF} in $d=5$ dimensions (blue) and free fermion \eqref{fermion} in $d=4$ dimensions (orange) -- in both cases $\Delta=3/2$ and $b_n$ are very close to each other and overlap in the plot. Also, Lanczos coefficients for composite operator $\phi^2$ in $d=5$ dimensions (green) and composite operator $\bar \psi \psi$ in $d=4$ (red) -- in both cases $\Delta=3$ and $b_n$ are again very close to each other and overlap.}
\label{fig:psiphi}
\end{figure}

Integration over the Matsubara propagator yields \eqref{fermion} with 
\bea
r_d=\beta^{1-d} {\Gamma\big({d\over 2}\big)  \over (4\pi)^{d\over 2}}.
\eea
It is valid only for $-1/2< \tau/\beta<1/2$ and should be extended by antiperiodicity
\bea
C_\psi(\tau+\beta)=-C_\psi(\tau)
\eea
beyond that. 
To evaluate $b_n^2$ numerically it is helpful to know  closed-form expression for the $2n$-th derivative 
\bea
 C_\psi^{(2n)}&=&r_d {\Gamma(2\Delta+2n)\over  2^{2n-1}\Gamma(2\Delta)}
 \big( \zeta\big(2\Delta+2n,{1\over4}\big) - \zeta\big(2\Delta +2n,{3\over4}\big) \big).
 \nonumber \\
\eea
As is pointed out in the main text resulting Lanczos coefficients are numerically very close to those for free scalar of the same conformal dimension $\Delta$. The same applies for composite operators $\bar \psi \psi$ and $\phi^2$. We illustrate that by the plot in Fig.~\ref{fig:psiphi}.

\subsection{Holographic thermal 2pt function}
\begin{figure}[t]
\includegraphics[width=0.5\textwidth]{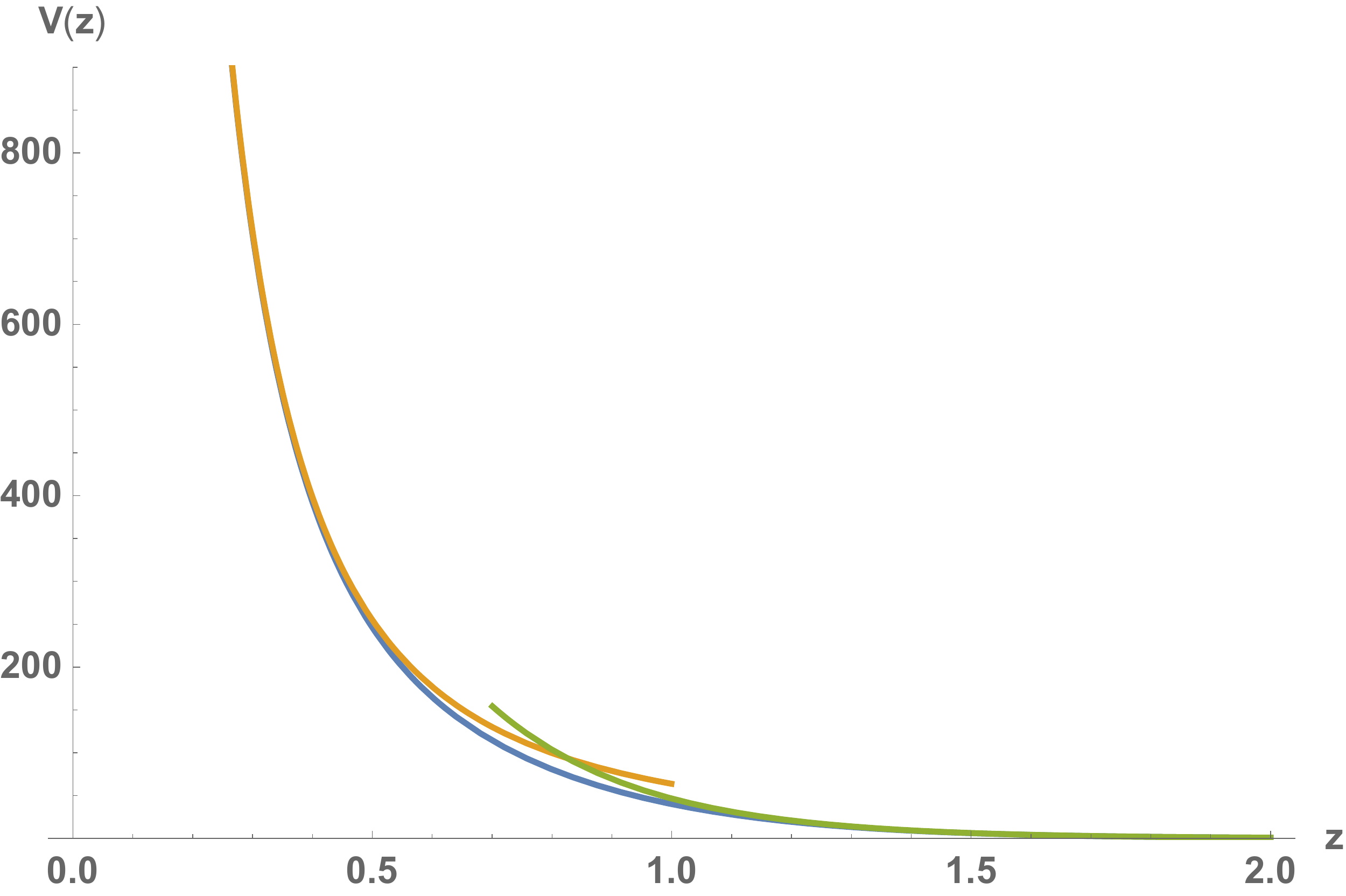}
\caption{Effective potential $V(z)$ \eqref{V} for $d=4,\nu=8$ (blue) vs asymptotic expressions \eqref{V1} (orange) and \eqref{V2} (green).}
\label{fig:V}
\end{figure}
\begin{figure}[t]
\includegraphics[width=0.5\textwidth]{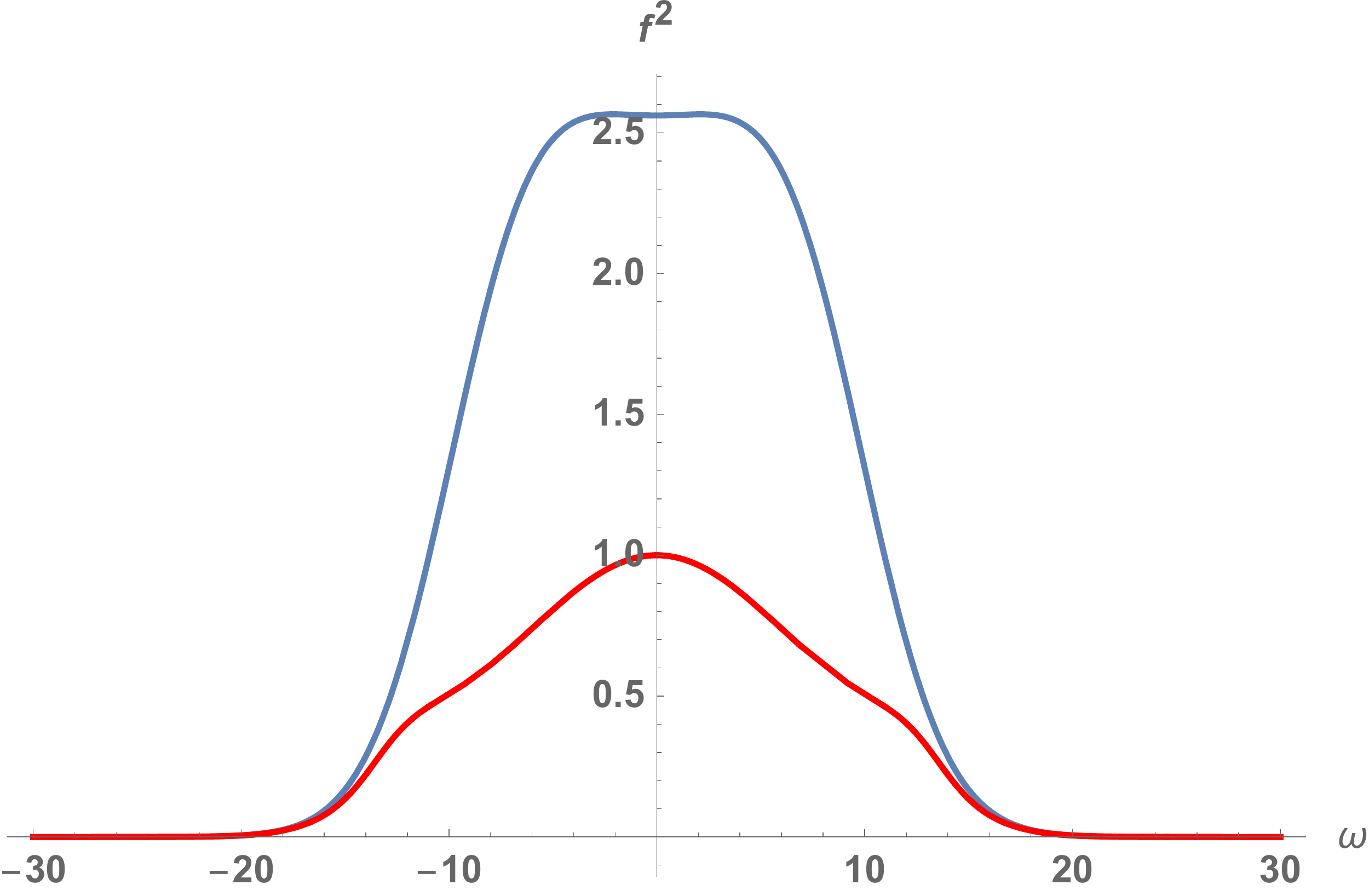}
\caption{Power spectrum $f^2$  given by \eqref{f2} for $d=4, \nu=8$ calculated using numerical integration from $z_1$ to $z_2$ (blue) and crude approximation by gluing $\psi_1$ and $\psi_2$ at $z=z^*$ (red). Both expression are multiplied by the same overall coefficient such that in the crude approximation case $f^2(0)=1$.  }
\label{fig:f2}
\end{figure}
\begin{figure}[b]
\includegraphics[width=0.5\textwidth]{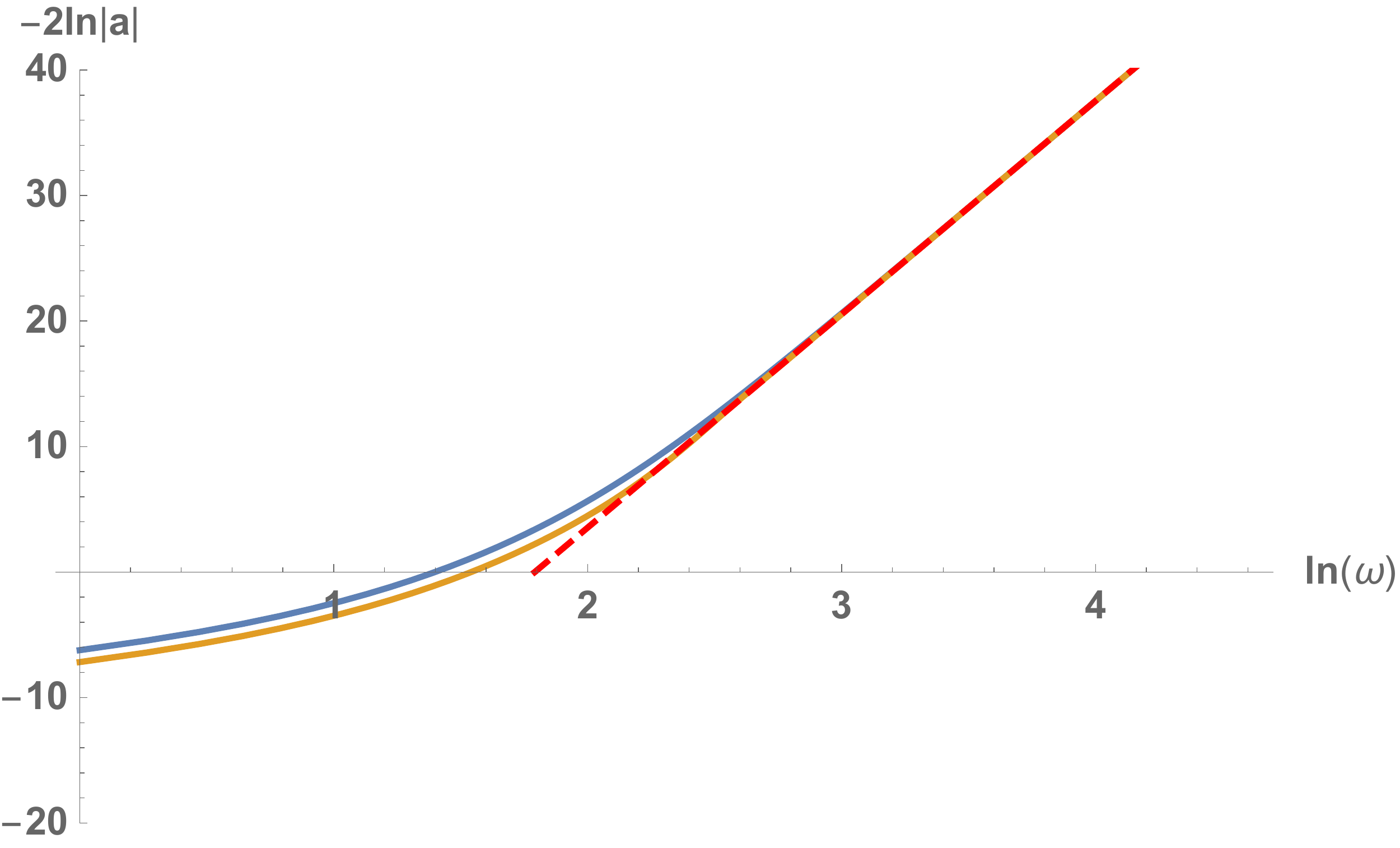}
\caption{Plot of $-2\ln|a|$ vs $\ln\omega$ calculated using  numerical integration from $z_1$ to $z_2$ (blue)  and crude approximation by gluing $\psi_1$ and $\psi_2$ at $z=z^*$ (orange) vs linear fit describing large $\omega$ behavior (red dashed line).}
\label{fig:c2}
\end{figure}
CFT on $\R^{1,d}$ at finite temperature $\beta$ is holographically described by a black brane background. To evaluate Wightman 2pt  function  $C_0(\tau)$ one needs to solve the wave-equation for a scalar field in the bulk dual to an operator  $\Op$ of dimension $\Delta$. For simplicity we will consider $\Op$ at zero spatial momentum, i.e.~our operator  in question is
\bea
O(t)={1\over \sqrt{\rm Volume}}\int d^{d-1} x\, \Op(t,x),
\eea
where overall normalization is chosen such that two-point function of $O$ is finite. 
Then the power spectrum $f^2(\omega)$ associated with 
\bea
\nonumber
C_0(\tau)&=&\Tr(e^{-(\beta-\tau) H/2} O e^{-(\beta+\tau)} O)=\\ && \int d^{d-1}x\, \langle \Op(i(\tau+\beta/2,\vec{x})\Op(0,\vec{0})\rangle_\beta,
\label{2pth}
\eea
is given by the following procedure \cite{festuccia2006excursions,festuccia2009bohr}. One introduces tortoise coordinate in the bulk 
\bea
z=\int_r^\infty {dr\over {\mathfrak f} },\qquad {\mathfrak f}=r^2-{1\over r^{d-2}},
\eea
and an effective potential 
\bea
\label{V}
V(z)={\mathfrak f}(r)\left(\nu^2-1/4+{(d-1)^2\over 4 r^d}\right),\\ 
\Delta=d/2+\nu. \nonumber
\eea
The scalar in the bulk dual to $\Op$ satisfies ``Schrodinger'' equation 
\bea
\label{SE}
-{d^2\psi\over dz^2}+V \psi=\omega^2 \psi,\quad 0\leq z<\infty.
\eea
Near $z\rightarrow 0$ the potential behaves as 
\bea
V(z)\approx {\nu^2-1/4\over z^2}, \label{V1}
\eea
and at large $z$
\bea
V(z)\approx  8 \left(\nu ^2+2\right) e^{\frac{\pi }{2}-4 z}, \label{V2}
\eea
where we restricted to $d=4$. The boundary behavior of $\psi$ is therefore 
\bea
\label{b1}
\psi(z)&\approx& z^{\nu+1/2},\quad z\rightarrow 0,\\
\psi(z)&\approx& a\, e^{-i\omega z}+\bar{a}\, e^{i\omega z},\quad z\rightarrow \infty,
\label{b2}
\eea
where $a$ is a complex $\omega$-dependent constant. The power spectrum of $C_0$ is then given by 
\bea
f^2(\omega) \propto {|a|^{-2}\over \omega \sinh(\beta \omega/2)}, \label{f2}
\eea
where temperature is fixed to be $\beta=4\pi/d$.
\begin{figure}[t]
\includegraphics[width=0.5\textwidth]{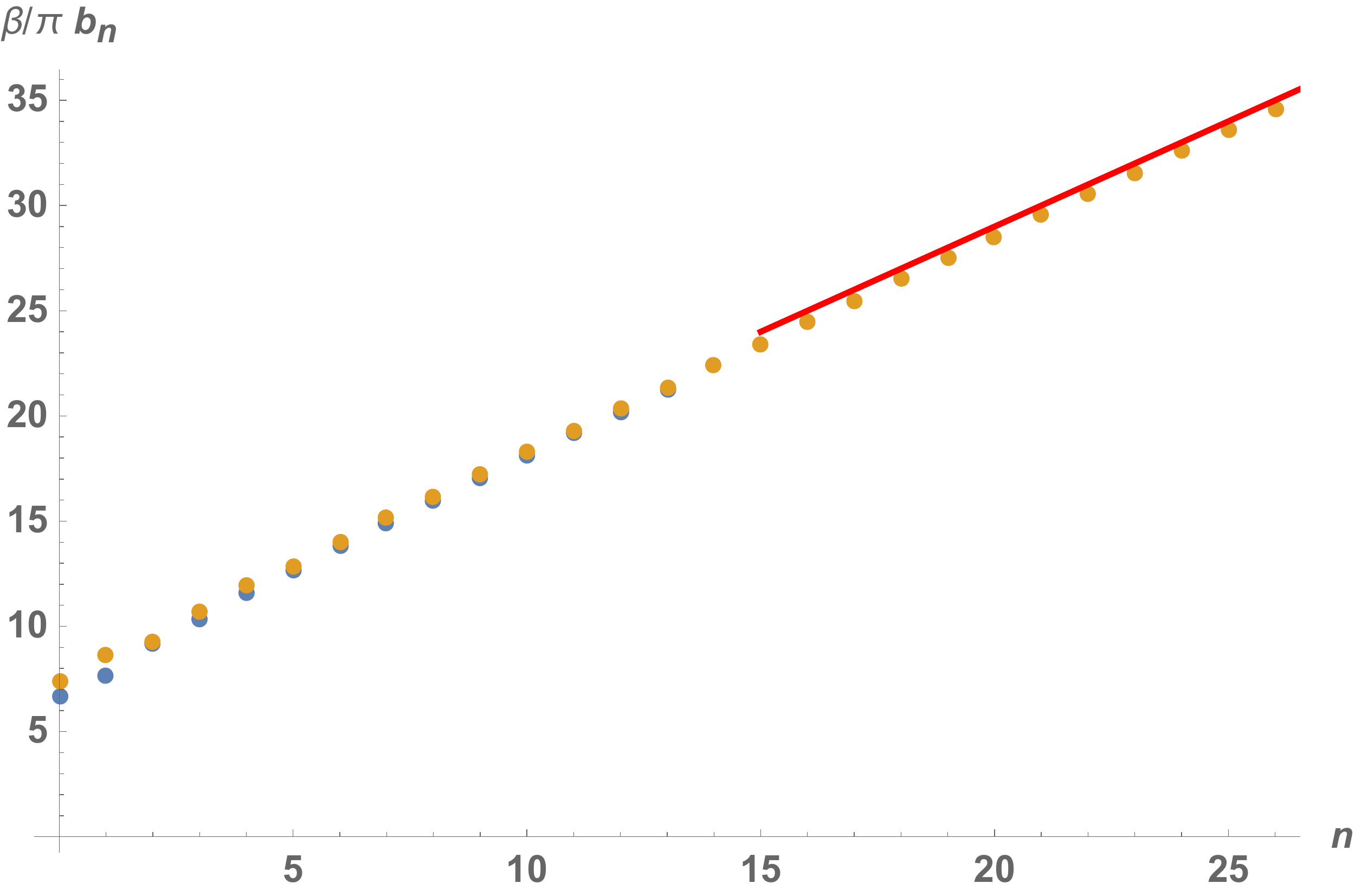}
\caption{``Holographic'' Lanczos coefficients $b_n$ for $0\leq n\leq 13$ evaluated using numerical integration from $z_1$ to $z_2$ (blue), $b_n$ evaluated using gluing $\psi_1$ and $\psi_2$ at $z=z^*$ (orange) and asymptotic behavior given by  \eqref{asympt}, $(\beta/\pi)b_n \approx n+\Delta -1$ (red line).}
\label{fig:b2nhol}
\end{figure}

To obtain $f^2(\omega)$ numerically one needs to solve \eqref{SE} with the boundary conditions (\ref{b1},\ref{b2}). In practice the asymptotic approximations (\ref{V1},\ref{V2}) accurately describe $V(z)$ everywhere outside of a small region of $z\sim 1$, as is shown in Fig.~\ref{fig:V} for $d=4$ and $\nu=8$ which corresponds to $\Delta=10$.
``Schrodinger'' equation \eqref{SE} with the approximate potential \eqref{V1} or \eqref{V2} can be solved analytically
\bea
\psi_1=\sqrt{z} \left(\frac{2}{\omega }\right)^{\nu } \Gamma (\nu +1) J_{\nu }(z \omega),
\eea
for small $z$ and 
\bea
\nonumber
\psi_2=a\, \Gamma \left(\frac{i \omega }{2}+1\right) I_{i\frac{\omega }{2}}\left(e^{\pi /4}\sqrt{2(\nu ^2+2)}\, e^{-2 z}\right)+{\rm c.c.}
\eea
for large $z$.  A crude approximation would be to neglect the region around $z\sim 1$ where asymptotic expressions \eqref{V1} and \eqref{V2} are less accurate and simply glue 
$\psi_1$ and $\psi_2$ at some intermediate point $z=z^*$ by continuity (and continuity of $\psi'$). We choose $z^*\approx 0.8251$ such that $V_1(z^*)=V_2(z^*)$. 
A more accurate approach would be to use $\psi_1$ for $z\leq z_1$ and $\psi_2$ for $z\geq z_2$, while integrating \eqref{SE} numerically from $z_1$ to $z_2$. 
We choose $z_1=0.2$ and $z_2=3$. Resulting profiles of $f^2(\omega)$ differ, as shown in Fig.~\ref{fig:f2}, but asymptotic behavior at large $\omega$ is the same. 
We confirm that by plotting $|a|^{-2}$ in logarithmic scale superimposed with a linear fit in Fig.~\ref{fig:c2}. 
The slope of the linear fit is $16.99$ which perfectly matches  the expected asymptotic behavior of $f^2$ \eqref{asymptotef2} after we take into account that 
\eqref{2pth} in the limit $\tau\rightarrow \beta/2$ has a singularity $C_0\propto (\tau-\beta/2)^{2\Delta-3}$.

With this we proceed to evaluate the moments \eqref{M2k} and Lanczos coefficients. The latter are shown in Fig.~\ref{fig:b2nhol} for both approximations superimposed with the asymptotic fit \eqref{asympt}.


\newpage
\bibliography{letter}

\begin{thebibliography}{39}%
\makeatletter
\providecommand \@ifxundefined [1]{%
 \@ifx{#1\undefined}
}%
\providecommand \@ifnum [1]{%
 \ifnum #1\expandafter \@firstoftwo
 \else \expandafter \@secondoftwo
 \fi
}%
\providecommand \@ifx [1]{%
 \ifx #1\expandafter \@firstoftwo
 \else \expandafter \@secondoftwo
 \fi
}%
\providecommand \natexlab [1]{#1}%
\providecommand \enquote  [1]{``#1''}%
\providecommand \bibnamefont  [1]{#1}%
\providecommand \bibfnamefont [1]{#1}%
\providecommand \citenamefont [1]{#1}%
\providecommand \href@noop [0]{\@secondoftwo}%
\providecommand \href [0]{\begingroup \@sanitize@url \@href}%
\providecommand \@href[1]{\@@startlink{#1}\@@href}%
\providecommand \@@href[1]{\endgroup#1\@@endlink}%
\providecommand \@sanitize@url [0]{\catcode `\\12\catcode `\$12\catcode
  `\&12\catcode `\#12\catcode `\^12\catcode `\_12\catcode `\%12\relax}%
\providecommand \@@startlink[1]{}%
\providecommand \@@endlink[0]{}%
\providecommand \url  [0]{\begingroup\@sanitize@url \@url }%
\providecommand \@url [1]{\endgroup\@href {#1}{\urlprefix }}%
\providecommand \urlprefix  [0]{URL }%
\providecommand \Eprint [0]{\href }%
\providecommand \doibase [0]{http://dx.doi.org/}%
\providecommand \selectlanguage [0]{\@gobble}%
\providecommand \bibinfo  [0]{\@secondoftwo}%
\providecommand \bibfield  [0]{\@secondoftwo}%
\providecommand \translation [1]{[#1]}%
\providecommand \BibitemOpen [0]{}%
\providecommand \bibitemStop [0]{}%
\providecommand \bibitemNoStop [0]{.\EOS\space}%
\providecommand \EOS [0]{\spacefactor3000\relax}%
\providecommand \BibitemShut  [1]{\csname bibitem#1\endcsname}%
\let\auto@bib@innerbib\@empty
\bibitem [{\citenamefont {D'Alessio}\ \emph {et~al.}(2016)\citenamefont
  {D'Alessio}, \citenamefont {Kafri}, \citenamefont {Polkovnikov},\ and\
  \citenamefont {Rigol}}]{d2016quantum}%
  \BibitemOpen
  \bibfield  {author} {\bibinfo {author} {\bibfnamefont {Luca}\ \bibnamefont
  {D'Alessio}}, \bibinfo {author} {\bibfnamefont {Yariv}\ \bibnamefont
  {Kafri}}, \bibinfo {author} {\bibfnamefont {Anatoli}\ \bibnamefont
  {Polkovnikov}}, \ and\ \bibinfo {author} {\bibfnamefont {Marcos}\
  \bibnamefont {Rigol}},\ }\bibfield  {title} {\enquote {\bibinfo {title} {From
  quantum chaos and eigenstate thermalization to statistical mechanics and
  thermodynamics},}\ }\href@noop {} {\bibfield  {journal} {\bibinfo  {journal}
  {Advances in Physics}\ }\textbf {\bibinfo {volume} {65}},\ \bibinfo {pages}
  {239--362} (\bibinfo {year} {2016})}\BibitemShut {NoStop}%
\bibitem [{\citenamefont {Maldacena}\ \emph {et~al.}(2016)\citenamefont
  {Maldacena}, \citenamefont {Shenker},\ and\ \citenamefont
  {Stanford}}]{maldacena2016bound}%
  \BibitemOpen
  \bibfield  {author} {\bibinfo {author} {\bibfnamefont {Juan}\ \bibnamefont
  {Maldacena}}, \bibinfo {author} {\bibfnamefont {Stephen~H}\ \bibnamefont
  {Shenker}}, \ and\ \bibinfo {author} {\bibfnamefont {Douglas}\ \bibnamefont
  {Stanford}},\ }\bibfield  {title} {\enquote {\bibinfo {title} {A bound on
  chaos},}\ }\href@noop {} {\bibfield  {journal} {\bibinfo  {journal} {Journal
  of High Energy Physics}\ }\textbf {\bibinfo {volume} {2016}},\ \bibinfo
  {pages} {1--17} (\bibinfo {year} {2016})}\BibitemShut {NoStop}%
\bibitem [{\citenamefont {Parker}\ \emph {et~al.}(2019)\citenamefont {Parker},
  \citenamefont {Cao}, \citenamefont {Avdoshkin}, \citenamefont {Scaffidi},\
  and\ \citenamefont {Altman}}]{parker2019universal}%
  \BibitemOpen
  \bibfield  {author} {\bibinfo {author} {\bibfnamefont {Daniel~E}\
  \bibnamefont {Parker}}, \bibinfo {author} {\bibfnamefont {Xiangyu}\
  \bibnamefont {Cao}}, \bibinfo {author} {\bibfnamefont {Alexander}\
  \bibnamefont {Avdoshkin}}, \bibinfo {author} {\bibfnamefont {Thomas}\
  \bibnamefont {Scaffidi}}, \ and\ \bibinfo {author} {\bibfnamefont {Ehud}\
  \bibnamefont {Altman}},\ }\bibfield  {title} {\enquote {\bibinfo {title} {A
  universal operator growth hypothesis},}\ }\href@noop {} {\bibfield  {journal}
  {\bibinfo  {journal} {Physical Review X}\ }\textbf {\bibinfo {volume} {9}},\
  \bibinfo {pages} {041017} (\bibinfo {year} {2019})}\BibitemShut {NoStop}%
\bibitem [{\citenamefont {Maldacena}\ and\ \citenamefont
  {Stanford}(2016)}]{PhysRevD.94.106002}%
  \BibitemOpen
  \bibfield  {author} {\bibinfo {author} {\bibfnamefont {Juan}\ \bibnamefont
  {Maldacena}}\ and\ \bibinfo {author} {\bibfnamefont {Douglas}\ \bibnamefont
  {Stanford}},\ }\bibfield  {title} {\enquote {\bibinfo {title} {Remarks on the
  sachdev-ye-kitaev model},}\ }\href {\doibase 10.1103/PhysRevD.94.106002}
  {\bibfield  {journal} {\bibinfo  {journal} {Phys. Rev. D}\ }\textbf {\bibinfo
  {volume} {94}},\ \bibinfo {pages} {106002} (\bibinfo {year}
  {2016})}\BibitemShut {NoStop}%
\bibitem [{\citenamefont {Rosenhaus}(2019)}]{rosenhaus2019introduction}%
  \BibitemOpen
  \bibfield  {author} {\bibinfo {author} {\bibfnamefont {Vladimir}\
  \bibnamefont {Rosenhaus}},\ }\bibfield  {title} {\enquote {\bibinfo {title}
  {An introduction to the syk model},}\ }\href@noop {} {\bibfield  {journal}
  {\bibinfo  {journal} {Journal of Physics A: Mathematical and Theoretical}\
  }\textbf {\bibinfo {volume} {52}},\ \bibinfo {pages} {323001} (\bibinfo
  {year} {2019})}\BibitemShut {NoStop}%
\bibitem [{\citenamefont {Trunin}(2020)}]{trunin2020pedagogical}%
  \BibitemOpen
  \bibfield  {author} {\bibinfo {author} {\bibfnamefont {Dmitrii~A}\
  \bibnamefont {Trunin}},\ }\bibfield  {title} {\enquote {\bibinfo {title}
  {Pedagogical introduction to syk model and 2d dilaton gravity},}\ }\href@noop
  {} {\bibfield  {journal} {\bibinfo  {journal} {arXiv preprint
  arXiv:2002.12187}\ } (\bibinfo {year} {2020})}\BibitemShut {NoStop}%
\bibitem [{\citenamefont {Roberts}\ \emph {et~al.}(2015)\citenamefont
  {Roberts}, \citenamefont {Stanford},\ and\ \citenamefont
  {Susskind}}]{roberts2015localized}%
  \BibitemOpen
  \bibfield  {author} {\bibinfo {author} {\bibfnamefont {Daniel~A}\
  \bibnamefont {Roberts}}, \bibinfo {author} {\bibfnamefont {Douglas}\
  \bibnamefont {Stanford}}, \ and\ \bibinfo {author} {\bibfnamefont {Leonard}\
  \bibnamefont {Susskind}},\ }\bibfield  {title} {\enquote {\bibinfo {title}
  {Localized shocks},}\ }\href@noop {} {\bibfield  {journal} {\bibinfo
  {journal} {Journal of High Energy Physics}\ }\textbf {\bibinfo {volume}
  {2015}},\ \bibinfo {pages} {51} (\bibinfo {year} {2015})}\BibitemShut
  {NoStop}%
\bibitem [{\citenamefont {Hartman}\ \emph {et~al.}(2017)\citenamefont
  {Hartman}, \citenamefont {Hartnoll},\ and\ \citenamefont
  {Mahajan}}]{hartman2017upper}%
  \BibitemOpen
  \bibfield  {author} {\bibinfo {author} {\bibfnamefont {Thomas}\ \bibnamefont
  {Hartman}}, \bibinfo {author} {\bibfnamefont {Sean~A}\ \bibnamefont
  {Hartnoll}}, \ and\ \bibinfo {author} {\bibfnamefont {Raghu}\ \bibnamefont
  {Mahajan}},\ }\bibfield  {title} {\enquote {\bibinfo {title} {Upper bound on
  diffusivity},}\ }\href@noop {} {\bibfield  {journal} {\bibinfo  {journal}
  {Physical review letters}\ }\textbf {\bibinfo {volume} {119}},\ \bibinfo
  {pages} {141601} (\bibinfo {year} {2017})}\BibitemShut {NoStop}%
\bibitem [{\citenamefont {Murthy}\ and\ \citenamefont
  {Srednicki}(2019)}]{murthy2019bounds}%
  \BibitemOpen
  \bibfield  {author} {\bibinfo {author} {\bibfnamefont {Chaitanya}\
  \bibnamefont {Murthy}}\ and\ \bibinfo {author} {\bibfnamefont {Mark}\
  \bibnamefont {Srednicki}},\ }\bibfield  {title} {\enquote {\bibinfo {title}
  {Bounds on chaos from the eigenstate thermalization hypothesis},}\
  }\href@noop {} {\bibfield  {journal} {\bibinfo  {journal} {Physical review
  letters}\ }\textbf {\bibinfo {volume} {123}},\ \bibinfo {pages} {230606}
  (\bibinfo {year} {2019})}\BibitemShut {NoStop}%
\bibitem [{\citenamefont {Barb{\'o}n}\ \emph {et~al.}(2019)\citenamefont
  {Barb{\'o}n}, \citenamefont {Rabinovici}, \citenamefont {Shir},\ and\
  \citenamefont {Sinha}}]{barbon2019evolution}%
  \BibitemOpen
  \bibfield  {author} {\bibinfo {author} {\bibfnamefont {JLF}\ \bibnamefont
  {Barb{\'o}n}}, \bibinfo {author} {\bibfnamefont {E}~\bibnamefont
  {Rabinovici}}, \bibinfo {author} {\bibfnamefont {R}~\bibnamefont {Shir}}, \
  and\ \bibinfo {author} {\bibfnamefont {R}~\bibnamefont {Sinha}},\ }\bibfield
  {title} {\enquote {\bibinfo {title} {On the evolution of operator complexity
  beyond scrambling},}\ }\href@noop {} {\bibfield  {journal} {\bibinfo
  {journal} {Journal of High Energy Physics}\ }\textbf {\bibinfo {volume}
  {2019}},\ \bibinfo {pages} {1--25} (\bibinfo {year} {2019})}\BibitemShut
  {NoStop}%
\bibitem [{\citenamefont {Jian}\ \emph {et~al.}(2020)\citenamefont {Jian},
  \citenamefont {Swingle},\ and\ \citenamefont {Xian}}]{jian2020complexity}%
  \BibitemOpen
  \bibfield  {author} {\bibinfo {author} {\bibfnamefont {Shao-Kai}\
  \bibnamefont {Jian}}, \bibinfo {author} {\bibfnamefont {Brian}\ \bibnamefont
  {Swingle}}, \ and\ \bibinfo {author} {\bibfnamefont {Zhuo-Yu}\ \bibnamefont
  {Xian}},\ }\bibfield  {title} {\enquote {\bibinfo {title} {Complexity growth
  of operators in the syk model and in jt gravity},}\ }\href@noop {} {\bibfield
   {journal} {\bibinfo  {journal} {arXiv preprint arXiv:2008.12274}\ }
  (\bibinfo {year} {2020})}\BibitemShut {NoStop}%
\bibitem [{\citenamefont {Rabinovici}\ \emph {et~al.}(2020)\citenamefont
  {Rabinovici}, \citenamefont {S{\'a}nchez-Garrido}, \citenamefont {Shir},\
  and\ \citenamefont {Sonner}}]{rabinovici2020operator}%
  \BibitemOpen
  \bibfield  {author} {\bibinfo {author} {\bibfnamefont {E}~\bibnamefont
  {Rabinovici}}, \bibinfo {author} {\bibfnamefont {A}~\bibnamefont
  {S{\'a}nchez-Garrido}}, \bibinfo {author} {\bibfnamefont {R}~\bibnamefont
  {Shir}}, \ and\ \bibinfo {author} {\bibfnamefont {J}~\bibnamefont {Sonner}},\
  }\bibfield  {title} {\enquote {\bibinfo {title} {Operator complexity: a
  journey to the edge of krylov space},}\ }\href@noop {} {\bibfield  {journal}
  {\bibinfo  {journal} {arXiv preprint arXiv:2009.01862}\ } (\bibinfo {year}
  {2020})}\BibitemShut {NoStop}%
\bibitem [{\citenamefont {Brown}\ \emph {et~al.}(2016)\citenamefont {Brown},
  \citenamefont {Roberts}, \citenamefont {Susskind}, \citenamefont {Swingle},\
  and\ \citenamefont {Zhao}}]{brown2016holographic}%
  \BibitemOpen
  \bibfield  {author} {\bibinfo {author} {\bibfnamefont {Adam~R}\ \bibnamefont
  {Brown}}, \bibinfo {author} {\bibfnamefont {Daniel~A}\ \bibnamefont
  {Roberts}}, \bibinfo {author} {\bibfnamefont {Leonard}\ \bibnamefont
  {Susskind}}, \bibinfo {author} {\bibfnamefont {Brian}\ \bibnamefont
  {Swingle}}, \ and\ \bibinfo {author} {\bibfnamefont {Ying}\ \bibnamefont
  {Zhao}},\ }\bibfield  {title} {\enquote {\bibinfo {title} {Holographic
  complexity equals bulk action?}}\ }\href@noop {} {\bibfield  {journal}
  {\bibinfo  {journal} {Physical review letters}\ }\textbf {\bibinfo {volume}
  {116}},\ \bibinfo {pages} {191301} (\bibinfo {year} {2016})}\BibitemShut
  {NoStop}%
\bibitem [{\citenamefont {Nunez}\ and\ \citenamefont
  {Starinets}(2003)}]{nunez2003ads}%
  \BibitemOpen
  \bibfield  {author} {\bibinfo {author} {\bibfnamefont {Alvaro}\ \bibnamefont
  {Nunez}}\ and\ \bibinfo {author} {\bibfnamefont {Andrei~O}\ \bibnamefont
  {Starinets}},\ }\bibfield  {title} {\enquote {\bibinfo {title} {Ads/cft
  correspondence, quasinormal modes, and thermal correlators in n= 4
  supersymmetric yang-mills theory},}\ }\href@noop {} {\bibfield  {journal}
  {\bibinfo  {journal} {Physical Review D}\ }\textbf {\bibinfo {volume} {67}},\
  \bibinfo {pages} {124013} (\bibinfo {year} {2003})}\BibitemShut {NoStop}%
\bibitem [{\citenamefont {Fidkowski}\ \emph {et~al.}(2004)\citenamefont
  {Fidkowski}, \citenamefont {Hubeny}, \citenamefont {Kleban},\ and\
  \citenamefont {Shenker}}]{fidkowski2004black}%
  \BibitemOpen
  \bibfield  {author} {\bibinfo {author} {\bibfnamefont {Lukasz}\ \bibnamefont
  {Fidkowski}}, \bibinfo {author} {\bibfnamefont {Veronika}\ \bibnamefont
  {Hubeny}}, \bibinfo {author} {\bibfnamefont {Matthew}\ \bibnamefont
  {Kleban}}, \ and\ \bibinfo {author} {\bibfnamefont {Stephen}\ \bibnamefont
  {Shenker}},\ }\bibfield  {title} {\enquote {\bibinfo {title} {The black hole
  singularity in ads/cft},}\ }\href@noop {} {\bibfield  {journal} {\bibinfo
  {journal} {Journal of High Energy Physics}\ }\textbf {\bibinfo {volume}
  {2004}},\ \bibinfo {pages} {014} (\bibinfo {year} {2004})}\BibitemShut
  {NoStop}%
\bibitem [{\citenamefont {Festuccia}\ and\ \citenamefont
  {Liu}(2006)}]{festuccia2006excursions}%
  \BibitemOpen
  \bibfield  {author} {\bibinfo {author} {\bibfnamefont {Guido}\ \bibnamefont
  {Festuccia}}\ and\ \bibinfo {author} {\bibfnamefont {Hong}\ \bibnamefont
  {Liu}},\ }\bibfield  {title} {\enquote {\bibinfo {title} {Excursions beyond
  the horizon: Black hole singularities in yang-mills theories (i)},}\
  }\href@noop {} {\bibfield  {journal} {\bibinfo  {journal} {Journal of High
  Energy Physics}\ }\textbf {\bibinfo {volume} {2006}},\ \bibinfo {pages} {044}
  (\bibinfo {year} {2006})}\BibitemShut {NoStop}%
\bibitem [{\citenamefont {Festuccia}\ and\ \citenamefont
  {Liu}(2009)}]{festuccia2009bohr}%
  \BibitemOpen
  \bibfield  {author} {\bibinfo {author} {\bibfnamefont {Guido}\ \bibnamefont
  {Festuccia}}\ and\ \bibinfo {author} {\bibfnamefont {Hong}\ \bibnamefont
  {Liu}},\ }\bibfield  {title} {\enquote {\bibinfo {title} {A bohr-sommerfeld
  quantization formula for quasinormal frequencies of ads black holes},}\
  }\href@noop {} {\bibfield  {journal} {\bibinfo  {journal} {Advanced Science
  Letters}\ }\textbf {\bibinfo {volume} {2}},\ \bibinfo {pages} {221--235}
  (\bibinfo {year} {2009})}\BibitemShut {NoStop}%
\bibitem [{\citenamefont {Iliesiu}\ \emph {et~al.}(2018)\citenamefont
  {Iliesiu}, \citenamefont {Kolo{\u{g}}lu}, \citenamefont {Mahajan},
  \citenamefont {Perlmutter},\ and\ \citenamefont
  {Simmons-Duffin}}]{iliesiu2018conformal}%
  \BibitemOpen
  \bibfield  {author} {\bibinfo {author} {\bibfnamefont {Luca}\ \bibnamefont
  {Iliesiu}}, \bibinfo {author} {\bibfnamefont {Murat}\ \bibnamefont
  {Kolo{\u{g}}lu}}, \bibinfo {author} {\bibfnamefont {Raghu}\ \bibnamefont
  {Mahajan}}, \bibinfo {author} {\bibfnamefont {Eric}\ \bibnamefont
  {Perlmutter}}, \ and\ \bibinfo {author} {\bibfnamefont {David}\ \bibnamefont
  {Simmons-Duffin}},\ }\bibfield  {title} {\enquote {\bibinfo {title} {The
  conformal bootstrap at finite temperature},}\ }\href@noop {} {\bibfield
  {journal} {\bibinfo  {journal} {Journal of High Energy Physics}\ }\textbf
  {\bibinfo {volume} {2018}},\ \bibinfo {pages} {1--71} (\bibinfo {year}
  {2018})}\BibitemShut {NoStop}%
\bibitem [{\citenamefont {Alday}\ \emph {et~al.}(2020)\citenamefont {Alday},
  \citenamefont {Kologlu},\ and\ \citenamefont
  {Zhiboedov}}]{alday2020holographic}%
  \BibitemOpen
  \bibfield  {author} {\bibinfo {author} {\bibfnamefont {Luis~F}\ \bibnamefont
  {Alday}}, \bibinfo {author} {\bibfnamefont {Murat}\ \bibnamefont {Kologlu}},
  \ and\ \bibinfo {author} {\bibfnamefont {Alexander}\ \bibnamefont
  {Zhiboedov}},\ }\bibfield  {title} {\enquote {\bibinfo {title} {Holographic
  correlators at finite temperature},}\ }\href@noop {} {\bibfield  {journal}
  {\bibinfo  {journal} {arXiv preprint arXiv:2009.10062}\ } (\bibinfo {year}
  {2020})}\BibitemShut {NoStop}%
\bibitem [{\citenamefont {Karlsson}\ \emph {et~al.}(2021)\citenamefont
  {Karlsson}, \citenamefont {Parnachev},\ and\ \citenamefont
  {Tadi{\'c}}}]{karlsson2021thermalization}%
  \BibitemOpen
  \bibfield  {author} {\bibinfo {author} {\bibfnamefont {Robin}\ \bibnamefont
  {Karlsson}}, \bibinfo {author} {\bibfnamefont {Andrei}\ \bibnamefont
  {Parnachev}}, \ and\ \bibinfo {author} {\bibfnamefont {Petar}\ \bibnamefont
  {Tadi{\'c}}},\ }\bibfield  {title} {\enquote {\bibinfo {title}
  {Thermalization in large-n cfts},}\ }\href@noop {} {\bibfield  {journal}
  {\bibinfo  {journal} {arXiv preprint arXiv:2102.04953}\ } (\bibinfo {year}
  {2021})}\BibitemShut {NoStop}%
\bibitem [{\citenamefont {Rodriguez-Gomez}\ and\ \citenamefont
  {Russo}(2021)}]{rodriguez2021correlation}%
  \BibitemOpen
  \bibfield  {author} {\bibinfo {author} {\bibfnamefont {D}~\bibnamefont
  {Rodriguez-Gomez}}\ and\ \bibinfo {author} {\bibfnamefont {JG}~\bibnamefont
  {Russo}},\ }\bibfield  {title} {\enquote {\bibinfo {title} {Correlation
  functions in finite temperature cft and black hole singularities},}\
  }\href@noop {} {\bibfield  {journal} {\bibinfo  {journal} {arXiv preprint
  arXiv:2102.11891}\ } (\bibinfo {year} {2021})}\BibitemShut {NoStop}%
\bibitem [{\citenamefont {Dymarsky}\ and\ \citenamefont
  {Gorsky}(2020)}]{dymarsky2020quantum}%
  \BibitemOpen
  \bibfield  {author} {\bibinfo {author} {\bibfnamefont {Anatoly}\ \bibnamefont
  {Dymarsky}}\ and\ \bibinfo {author} {\bibfnamefont {Alexander}\ \bibnamefont
  {Gorsky}},\ }\bibfield  {title} {\enquote {\bibinfo {title} {Quantum chaos as
  delocalization in krylov space},}\ }\href@noop {} {\bibfield  {journal}
  {\bibinfo  {journal} {Physical Review B}\ }\textbf {\bibinfo {volume}
  {102}},\ \bibinfo {pages} {085137} (\bibinfo {year} {2020})}\BibitemShut
  {NoStop}%
\bibitem [{\citenamefont {Elsayed}\ \emph {et~al.}(2014)\citenamefont
  {Elsayed}, \citenamefont {Hess},\ and\ \citenamefont
  {Fine}}]{elsayed2014signatures}%
  \BibitemOpen
  \bibfield  {author} {\bibinfo {author} {\bibfnamefont {Tarek~A}\ \bibnamefont
  {Elsayed}}, \bibinfo {author} {\bibfnamefont {Benjamin}\ \bibnamefont
  {Hess}}, \ and\ \bibinfo {author} {\bibfnamefont {Boris~V}\ \bibnamefont
  {Fine}},\ }\bibfield  {title} {\enquote {\bibinfo {title} {Signatures of
  chaos in time series generated by many-spin systems at high temperatures},}\
  }\href@noop {} {\bibfield  {journal} {\bibinfo  {journal} {Physical Review
  E}\ }\textbf {\bibinfo {volume} {90}},\ \bibinfo {pages} {022910} (\bibinfo
  {year} {2014})}\BibitemShut {NoStop}%
\bibitem [{\citenamefont {Avdoshkin}\ and\ \citenamefont
  {Dymarsky}(2020)}]{avdoshkin2020euclidean}%
  \BibitemOpen
  \bibfield  {author} {\bibinfo {author} {\bibfnamefont {Alexander}\
  \bibnamefont {Avdoshkin}}\ and\ \bibinfo {author} {\bibfnamefont {Anatoly}\
  \bibnamefont {Dymarsky}},\ }\bibfield  {title} {\enquote {\bibinfo {title}
  {Euclidean operator growth and quantum chaos},}\ }\href@noop {} {\bibfield
  {journal} {\bibinfo  {journal} {Physical Review Research}\ }\textbf {\bibinfo
  {volume} {2}},\ \bibinfo {pages} {043234} (\bibinfo {year}
  {2020})}\BibitemShut {NoStop}%
\bibitem [{\citenamefont {Lubinsky}(1993)}]{lubinsky1993update}%
  \BibitemOpen
  \bibfield  {author} {\bibinfo {author} {\bibfnamefont {DS}~\bibnamefont
  {Lubinsky}},\ }\bibfield  {title} {\enquote {\bibinfo {title} {An update on
  orthogonal polynomials and weighted approximation on the real line},}\
  }\href@noop {} {\bibfield  {journal} {\bibinfo  {journal} {Acta Applicandae
  Mathematica}\ }\textbf {\bibinfo {volume} {33}},\ \bibinfo {pages} {121--164}
  (\bibinfo {year} {1993})}\BibitemShut {NoStop}%
\bibitem [{\citenamefont {Basor}\ \emph {et~al.}(2001)\citenamefont {Basor},
  \citenamefont {Chen},\ and\ \citenamefont {Widom}}]{basor2001determinants}%
  \BibitemOpen
  \bibfield  {author} {\bibinfo {author} {\bibfnamefont {Estelle~L}\
  \bibnamefont {Basor}}, \bibinfo {author} {\bibfnamefont {Yang}\ \bibnamefont
  {Chen}}, \ and\ \bibinfo {author} {\bibfnamefont {Harold}\ \bibnamefont
  {Widom}},\ }\bibfield  {title} {\enquote {\bibinfo {title} {Determinants of
  hankel matrices},}\ }\href@noop {} {\bibfield  {journal} {\bibinfo  {journal}
  {Journal of Functional Analysis}\ }\textbf {\bibinfo {volume} {179}},\
  \bibinfo {pages} {214--234} (\bibinfo {year} {2001})}\BibitemShut {NoStop}%
\bibitem [{\citenamefont {Yates}\ \emph
  {et~al.}(2020{\natexlab{a}})\citenamefont {Yates}, \citenamefont {Abanov},\
  and\ \citenamefont {Mitra}}]{yates2020lifetime}%
  \BibitemOpen
  \bibfield  {author} {\bibinfo {author} {\bibfnamefont {Daniel~J}\
  \bibnamefont {Yates}}, \bibinfo {author} {\bibfnamefont {Alexander~G}\
  \bibnamefont {Abanov}}, \ and\ \bibinfo {author} {\bibfnamefont {Aditi}\
  \bibnamefont {Mitra}},\ }\bibfield  {title} {\enquote {\bibinfo {title}
  {Lifetime of almost strong edge-mode operators in one-dimensional,
  interacting, symmetry protected topological phases},}\ }\href@noop {}
  {\bibfield  {journal} {\bibinfo  {journal} {Physical Review Letters}\
  }\textbf {\bibinfo {volume} {124}},\ \bibinfo {pages} {206803} (\bibinfo
  {year} {2020}{\natexlab{a}})}\BibitemShut {NoStop}%
\bibitem [{\citenamefont {Yates}\ \emph
  {et~al.}(2020{\natexlab{b}})\citenamefont {Yates}, \citenamefont {Abanov},\
  and\ \citenamefont {Mitra}}]{yates2020dynamics}%
  \BibitemOpen
  \bibfield  {author} {\bibinfo {author} {\bibfnamefont {Daniel~J}\
  \bibnamefont {Yates}}, \bibinfo {author} {\bibfnamefont {Alexander~G}\
  \bibnamefont {Abanov}}, \ and\ \bibinfo {author} {\bibfnamefont {Aditi}\
  \bibnamefont {Mitra}},\ }\bibfield  {title} {\enquote {\bibinfo {title}
  {Dynamics of almost strong edge modes in spin chains away from
  integrability},}\ }\href@noop {} {\bibfield  {journal} {\bibinfo  {journal}
  {Physical Review B}\ }\textbf {\bibinfo {volume} {102}},\ \bibinfo {pages}
  {195419} (\bibinfo {year} {2020}{\natexlab{b}})}\BibitemShut {NoStop}%
\bibitem [{\citenamefont {Caputa}\ \emph {et~al.}(2016)\citenamefont {Caputa},
  \citenamefont {Numasawa},\ and\ \citenamefont
  {Veliz-Osorio}}]{caputa2016scrambling}%
  \BibitemOpen
  \bibfield  {author} {\bibinfo {author} {\bibfnamefont {Pawel}\ \bibnamefont
  {Caputa}}, \bibinfo {author} {\bibfnamefont {Tokiro}\ \bibnamefont
  {Numasawa}}, \ and\ \bibinfo {author} {\bibfnamefont {Alvaro}\ \bibnamefont
  {Veliz-Osorio}},\ }\bibfield  {title} {\enquote {\bibinfo {title} {Scrambling
  without chaos in rcft},}\ }\href@noop {} {\bibfield  {journal} {\bibinfo
  {journal} {arXiv preprint arXiv:1602.06542}\ } (\bibinfo {year}
  {2016})}\BibitemShut {NoStop}%
\bibitem [{\citenamefont {Fan}(2018)}]{fan2018out}%
  \BibitemOpen
  \bibfield  {author} {\bibinfo {author} {\bibfnamefont {Ruihua}\ \bibnamefont
  {Fan}},\ }\bibfield  {title} {\enquote {\bibinfo {title} {Out-of-time-order
  correlation functions for unitary minimal models},}\ }\href@noop {}
  {\bibfield  {journal} {\bibinfo  {journal} {arXiv preprint arXiv:1809.07228}\
  } (\bibinfo {year} {2018})}\BibitemShut {NoStop}%
\bibitem [{\citenamefont {Kudler-Flam}\ \emph {et~al.}(2020)\citenamefont
  {Kudler-Flam}, \citenamefont {Nie},\ and\ \citenamefont
  {Ryu}}]{kudler2020conformal}%
  \BibitemOpen
  \bibfield  {author} {\bibinfo {author} {\bibfnamefont {Jonah}\ \bibnamefont
  {Kudler-Flam}}, \bibinfo {author} {\bibfnamefont {Laimei}\ \bibnamefont
  {Nie}}, \ and\ \bibinfo {author} {\bibfnamefont {Shinsei}\ \bibnamefont
  {Ryu}},\ }\bibfield  {title} {\enquote {\bibinfo {title} {Conformal field
  theory and the web of quantum chaos diagnostics},}\ }\href@noop {} {\bibfield
   {journal} {\bibinfo  {journal} {Journal of High Energy Physics}\ }\textbf
  {\bibinfo {volume} {2020}},\ \bibinfo {pages} {1--33} (\bibinfo {year}
  {2020})}\BibitemShut {NoStop}%
\bibitem [{\citenamefont {Stanford}(2016)}]{stanford2016many}%
  \BibitemOpen
  \bibfield  {author} {\bibinfo {author} {\bibfnamefont {Douglas}\ \bibnamefont
  {Stanford}},\ }\bibfield  {title} {\enquote {\bibinfo {title} {Many-body
  chaos at weak coupling},}\ }\href@noop {} {\bibfield  {journal} {\bibinfo
  {journal} {Journal of High Energy Physics}\ }\textbf {\bibinfo {volume}
  {2016}},\ \bibinfo {pages} {1--18} (\bibinfo {year} {2016})}\BibitemShut
  {NoStop}%
\bibitem [{\citenamefont {Murugan}\ \emph {et~al.}(2017)\citenamefont
  {Murugan}, \citenamefont {Stanford},\ and\ \citenamefont
  {Witten}}]{murugan2017more}%
  \BibitemOpen
  \bibfield  {author} {\bibinfo {author} {\bibfnamefont {Jeff}\ \bibnamefont
  {Murugan}}, \bibinfo {author} {\bibfnamefont {Douglas}\ \bibnamefont
  {Stanford}}, \ and\ \bibinfo {author} {\bibfnamefont {Edward}\ \bibnamefont
  {Witten}},\ }\bibfield  {title} {\enquote {\bibinfo {title} {More on
  supersymmetric and 2d analogs of the syk model},}\ }\href@noop {} {\bibfield
  {journal} {\bibinfo  {journal} {Journal of High Energy Physics}\ }\textbf
  {\bibinfo {volume} {2017}},\ \bibinfo {pages} {1--99} (\bibinfo {year}
  {2017})}\BibitemShut {NoStop}%
\bibitem [{\citenamefont {Steinberg}\ and\ \citenamefont
  {Swingle}(2019)}]{steinberg2019thermalization}%
  \BibitemOpen
  \bibfield  {author} {\bibinfo {author} {\bibfnamefont {Julia}\ \bibnamefont
  {Steinberg}}\ and\ \bibinfo {author} {\bibfnamefont {Brian}\ \bibnamefont
  {Swingle}},\ }\bibfield  {title} {\enquote {\bibinfo {title} {Thermalization
  and chaos in qed 3},}\ }\href@noop {} {\bibfield  {journal} {\bibinfo
  {journal} {Physical Review D}\ }\textbf {\bibinfo {volume} {99}},\ \bibinfo
  {pages} {076007} (\bibinfo {year} {2019})}\BibitemShut {NoStop}%
\bibitem [{\citenamefont {Mezei}\ and\ \citenamefont
  {S{\'a}rosi}(2020)}]{mezei2020chaos}%
  \BibitemOpen
  \bibfield  {author} {\bibinfo {author} {\bibfnamefont {M{\'a}rk}\
  \bibnamefont {Mezei}}\ and\ \bibinfo {author} {\bibfnamefont {G{\'a}bor}\
  \bibnamefont {S{\'a}rosi}},\ }\bibfield  {title} {\enquote {\bibinfo {title}
  {Chaos in the butterfly cone},}\ }\href@noop {} {\bibfield  {journal}
  {\bibinfo  {journal} {Journal of High Energy Physics}\ }\textbf {\bibinfo
  {volume} {2020}},\ \bibinfo {pages} {1--34} (\bibinfo {year}
  {2020})}\BibitemShut {NoStop}%
\bibitem [{\citenamefont {Jefferson}\ and\ \citenamefont
  {Myers}(2017)}]{jefferson2017circuit}%
  \BibitemOpen
  \bibfield  {author} {\bibinfo {author} {\bibfnamefont {Robert~A}\
  \bibnamefont {Jefferson}}\ and\ \bibinfo {author} {\bibfnamefont {Robert~C}\
  \bibnamefont {Myers}},\ }\bibfield  {title} {\enquote {\bibinfo {title}
  {Circuit complexity in quantum field theory},}\ }\href@noop {} {\bibfield
  {journal} {\bibinfo  {journal} {Journal of High Energy Physics}\ }\textbf
  {\bibinfo {volume} {2017}},\ \bibinfo {pages} {1--81} (\bibinfo {year}
  {2017})}\BibitemShut {NoStop}%
\bibitem [{\citenamefont {Chapman}\ \emph {et~al.}(2018)\citenamefont
  {Chapman}, \citenamefont {Heller}, \citenamefont {Marrochio},\ and\
  \citenamefont {Pastawski}}]{PhysRevLett.120.121602}%
  \BibitemOpen
  \bibfield  {author} {\bibinfo {author} {\bibfnamefont {Shira}\ \bibnamefont
  {Chapman}}, \bibinfo {author} {\bibfnamefont {Michal~P.}\ \bibnamefont
  {Heller}}, \bibinfo {author} {\bibfnamefont {Hugo}\ \bibnamefont
  {Marrochio}}, \ and\ \bibinfo {author} {\bibfnamefont {Fernando}\
  \bibnamefont {Pastawski}},\ }\bibfield  {title} {\enquote {\bibinfo {title}
  {Toward a definition of complexity for quantum field theory states},}\ }\href
  {\doibase 10.1103/PhysRevLett.120.121602} {\bibfield  {journal} {\bibinfo
  {journal} {Phys. Rev. Lett.}\ }\textbf {\bibinfo {volume} {120}},\ \bibinfo
  {pages} {121602} (\bibinfo {year} {2018})}\BibitemShut {NoStop}%
\bibitem [{\citenamefont {Caputa}\ \emph {et~al.}(2017)\citenamefont {Caputa},
  \citenamefont {Kundu}, \citenamefont {Miyaji}, \citenamefont {Takayanagi},\
  and\ \citenamefont {Watanabe}}]{caputa2017liouville}%
  \BibitemOpen
  \bibfield  {author} {\bibinfo {author} {\bibfnamefont {Pawel}\ \bibnamefont
  {Caputa}}, \bibinfo {author} {\bibfnamefont {Nilay}\ \bibnamefont {Kundu}},
  \bibinfo {author} {\bibfnamefont {Masamichi}\ \bibnamefont {Miyaji}},
  \bibinfo {author} {\bibfnamefont {Tadashi}\ \bibnamefont {Takayanagi}}, \
  and\ \bibinfo {author} {\bibfnamefont {Kento}\ \bibnamefont {Watanabe}},\
  }\bibfield  {title} {\enquote {\bibinfo {title} {Liouville action as
  path-integral complexity: from continuous tensor networks to ads/cft},}\
  }\href@noop {} {\bibfield  {journal} {\bibinfo  {journal} {Journal of High
  Energy Physics}\ }\textbf {\bibinfo {volume} {2017}},\ \bibinfo {pages} {97}
  (\bibinfo {year} {2017})}\BibitemShut {NoStop}%
\bibitem [{\citenamefont {Caputa}\ and\ \citenamefont
  {Magan}(2019)}]{caputa2019quantum}%
  \BibitemOpen
  \bibfield  {author} {\bibinfo {author} {\bibfnamefont {Pawe{\l}}\
  \bibnamefont {Caputa}}\ and\ \bibinfo {author} {\bibfnamefont {Javier~M}\
  \bibnamefont {Magan}},\ }\bibfield  {title} {\enquote {\bibinfo {title}
  {Quantum computation as gravity},}\ }\href@noop {} {\bibfield  {journal}
  {\bibinfo  {journal} {Physical review letters}\ }\textbf {\bibinfo {volume}
  {122}},\ \bibinfo {pages} {231302} (\bibinfo {year} {2019})}\BibitemShut
  {NoStop}%
\end{thebibliography}%

\end{document}